%% file: rs_resum.tex
\documentclass[a4paper,11pt]{article}
\pdfoutput=1 
\usepackage{jheppub}  
\usepackage{graphicx,color}
\usepackage{amsmath}
\usepackage{autobreak}
\allowdisplaybreaks
\newcommand{\ben}{\begin{enumerate}}
\newcommand{\een}{\end{enumerate}}
\newcommand{\beq}{\begin{equation}}
\newcommand{\eeq}{\end{equation}}
\newcommand{\bal}{\begin{align}}
\newcommand{\eal}{\end{align}}

\newcommand{\bea}{\begin{eqnarray}}
\newcommand{\eea}{\end{eqnarray}}
\newcommand{\nn}{\nonumber}
\newcommand{\muf}{\mu_f}
\newcommand{\omg}{\omega}

\newcommand{\olsi}[1]{\,\overline{\!{#1}}} 
\newcommand{\Nb}{{\olsi{N}}}
\def\Dm1{{{\delta(1-z)}}}

\def\g0#1DY{{g_{0#1}^{DY}}}

\def\L{{\cal L}}

\def\LogmW1{{{\ln (1-\omega)}}}

\def\viz{{ {\it viz. }}}
\def\ie{{\it i.e. }}
\newcommand{\eq}[1]{eq.\ (\ref{#1})}
\newcommand{\fig}[1]{fig.\ (\ref{#1})}
\newcommand{\tab}[1]{tab.\ (\ref{#1})}
\newcommand{\sect}[1]{sec.\ (\ref{#1})}

\title{Resummed inclusive cross-section in Randall-Sundrum model at NNLO+NNLL}
\author[a]{Goutam Das,}
\author[b]{M. C. Kumar }
\author[b]{ and Kajal Samanta }
\affiliation[a]{Theoretische Physik 1, Naturwissenschaftlich-Technische Fakult{\"a}t, Universit{\"a}t Siegen,
Walter-Flex-Strasse 3, 57068 Siegen, Germany}
\affiliation[b]{Department of Physics, Indian Institute of Technology Guwahati, Guwahati-781039, Assam, India}
\emailAdd{goutam.das@uni-siegen.de}
\emailAdd{mckumar@iitg.ac.in}
\emailAdd{kajal.samanta@iitg.ac.in }
\abstract{
The complete next-to-next-to leading order (NNLO) QCD correction has been studied to the di-lepton invariant mass distribution within the Randall-Sundrum (RS) framework. In addition, the soft-virtual (SV) cross-section at next-to-next-to-next-to leading order (N$^3$LO) as well as threshold resummation to next-to-next-to leading logarithms (NNLL) level have been presented. The analytical coefficient for SV production has been obtained up to three loops very recently along with the process-dependent coefficients needed to perform resummation up to NNLL. These coefficients are universal for any universal spin-2 model where spin-2 particle couples to the Standard Model (SM) particles with equal strength. We use these coefficients in predicting N$^3$LO SV results as well as matched NNLO+NNLL results for invariant mass distribution for Drell-Yan (DY) production in RS model. We performed a detailed phenomenological analysis and present prediction for the 13 TeV centre-of-mass energy at the Large Hadron Collider (LHC) for the search of such RS Kaluza-Klein (KK) resonances. The NNLO cross-section adds about $21\%$  correction to the next-to-leading order (NLO) results. We found that the SV correction at the N$^3$LO order decreases the cross-section by $0.7\%$ near the first KK resonance ($M_1=1500$ GeV) whereas the resummed result shows an increment over NNLO by $7\%$ of LO. We performed a detailed analysis including scale variation and parton distribution function (PDF) variations. These new results provide an opportunity to stringently constrain the parameters of the model in particular in the search of heavy spin-2 resonances at the LHC.
}

\begin{document} 
\preprint{SI-HEP-2020-03}
\maketitle
\section{Introduction} \label{sec:introduction}

The Standard Model (SM) of particle physics is now well established after the discovery of scalar Higgs boson \cite{Aad:2012tfa,Chatrchyan:2012xdj}  at the Large Hadron Collider (LHC). The properties of the Higgs boson is being tested at a very high accuracy in the hope of new physics beyond the SM (BSM). A large class of the BSM scenarios are motivated by the large hierarchy between the electroweak symmetry breaking scale and the Planck scale, known as the gauge hierarchy problem. A wide class of theories have been proposed to address this problem through the introduction of large extra dimensions in the TeV scale brane world scenarios. In particular the models with warped extra dimension as proposed by Randall and Sundrum (RS) \cite{Randall:1999ee} are attractive candidates to solve this gauge hierarchy problem. In its simplest version, it predicts spin-2 Kaluza-Klein (KK) excitations in the TeV mass range which could be accessible at current hadron collider LHC or in any future hadron colliders or electron-positron colliders.

Search of physics beyond the SM has been an important objective of the LHC physics program. Precision physics plays an important role in this regard to accurately predict the cross-sections and distributions within perturbative framework. The process like Higgs and pseudo-scalar Higgs boson \cite{Harlander:2002wh,Harlander:2002vv,Anastasiou:2002yz,Anastasiou:2002wq,Ravindran:2003um,Harlander:2003ai}, DY \cite{Hamberg:1990np,Harlander:2002wh} productions are already available at NNLO accuracy. The large perturbative corrections for Higgs at NNLO even pushes the accuracy to be calculated to even N$^3$LO order  \cite{Anastasiou:2015ema,Mistlberger:2018etf,Duhr:2019kwi}. Recently the DY production has also been calculated to third order in strong coupling \cite{Duhr:2020seh}. The exclusive observables like rapidity are also being calculated to the same accuracy (see for example \cite{Anastasiou:2004xq,Anastasiou:2011qx,Buehler:2012cu,Dulat:2018bfe,Cieri:2018oms,Anastasiou:2003yy,Anastasiou:2003ds,Catani:2009sm,Melnikov:2006kv,Gavin:2012sy}).

In order to achieve perturbative stability, it is instructive to go beyond NNLO by computing the SV cross-section at N$^3$LO order. The SV corrections constitute a significant part of the full cross-section and have been successfully computed for many processes in the SM, for example, Higgs production \cite{Anastasiou:2014vaa,Moch:2005ky,Laenen:2005uz,Ravindran:2005vv,Ravindran:2006cg,Idilbi:2005ni, Li:2014afw, Ahmed:2014cha}, DY production \cite{Ravindran:2006bu,Ahmed:2014cla}, associated production  \cite{Kumar:2014uwa} to N$^3$LO as well as in BSM domain like pseudo-scalar production \cite{Ahmed:2015qda} in 2HDM. Similar accuracy has also been achieved for rapidity distributions \cite{Ravindran:2006bu,Ravindran:2007sv,Ahmed:2014uya,Ahmed:2014era}.

In the threshold region where partonic $z\to 1$, the truncated fixed order cross-section however becomes unreliable due to the presence of large logarithms. These large logarithms arise due to constrained phase space available for the soft gluons. In order to get a reliable prediction also in these corners of the phase-space, it is thus essential to resum these large logarithms to all orders.  Threshold resummation has been performed successfully to inclusive Higgs production \cite{Catani:2003zt,Moch:2005ky,Catani:2014uta,Bonvini:2014joa,Ahmed:2015sna,Bonvini:2016frm,H:2019dcl}, DY production \cite{Moch:2005ky,Catani:2014uta}, DIS \cite{Moch:2005ba} as well as for pseudo-scalar production \cite{Schmidt:2015cea,deFlorian:2007sr,Ahmed:2016otz} up to N$^3$LL accuracy. The first results towards N$^4$LL corrections are also available recently for DIS in \cite{Das:2019btv}. Moreover for differential observables like rapidity, it is known to NNLL accuracy for many important processes (see for example \cite{Catani:1989ne,Westmark:2017uig,Banerjee:2017cfc,Banerjee:2018vvb,Lustermans:2019cau,Ebert:2017uel}). 

In the context of large extra dimension, the NLO corrections  were known for many important processes at the LHC \cite{Mathews:2004xp,Kumar:2007af,Kumar:2008dn,Kumar:2008pk,Kumar:2009nn,Agarwal:2009xr,Agarwal:2010sp,Kumar:2011yta,Agarwal:2009zg,Agarwal:2010sn,Mathews:2005bw} within Arkani-Hamed-Dimopoulos-Dvali (ADD) \cite{ArkaniHamed:1998rs} and RS \cite{Randall:1999ee} model. It is observed in the NLO QCD computation \cite{Mathews:2004xp} that the K-factors in the di-lepton production case are potentially large and range up to 60\%. The matched NLO results with parton shower is also known for di-final processes in ADD  \cite{Frederix:2012dp,Frederix:2013lga} and in RS \cite{Das:2014tva} model. The associated production \cite{Kumar:2010kv} as well as triple gauge boson production processes \cite{Kumar:2011jq} are also known. In RS model, the triple neutral gauge boson productions are available \cite{Das:2015bda} in ME+PS accuracy in the M{\scriptsize AD}G{\scriptsize RAPH} framework. Moreover generic universal and non-universal spin-2 production  processes  are automatized \cite{Das:2016pbk} in F{{\scriptsize EYNRULES}} \cite{Alloul:2013bka} - M{\scriptsize AD}G{\scriptsize RAPH}5\_{\scriptsize A}MC@NLO \cite{Alwall:2014hca} framework   providing  NLO accuracy for inclusive and exclusive cross-sections for all relevant channels at the LHC.

The first attempt to go beyond the NLO accuracy has been seen in \cite{deFlorian:2013wpa} calculating the SV corrections at NNLO. This has been possible due to the calculation of spin-2 form factor \cite{deFlorian:2013sza} at the same order. Shortly after, the complete NNLO corrections were computed  in \cite{Ahmed:2016qhu} using the method of reverse unitarity \cite{Anastasiou:2002yz} and phenomenological study has been performed in the context of ADD model. There it has been found that the NNLO correction changes the cross-section by 21\% over NLO results and constrains the scale uncertainty to $1.6\%$. Similar accuracy is also available for non-universal spin-2 production \cite{Banerjee:2017ewt} where spin-2 couples with different coupling to the SM fields. The first attempt in calculating the SV corrections beyond NNLO can be seen \cite{Das:2019bxi} in the context of ADD model in DY invariant mass distribution after the completion of three-loop quark and gluon form factor \cite{Ahmed:2015qia}. The perturbative coefficients are same for any spin-2 production with universal coupling to the SM. There it has been noticed that the N$^3$LO SV cross-section changes the NNLO by -0.7\% at $Q = 1500$ GeV ($Q$ being the invariant mass of the lepton pair). Moreover the authors also performed threshold resummation up to N$^3$LL accuracy and the corrections are found to be around 1\% over NNLO with scale uncertainty reduces to $1.5\%$. 

In this article, we focus on massive KK production in the RS framework. Since the spin-2 RS  KK excitations also couples universally to the SM stress-energy tensor, the analytical perturbative coefficients are same as the generic universal spin-2 case like ADD. Phenomenologically however the RS KK states provide very distinctive signature from that of ADD model at the LHC. Where the di-lepton invariant mass distribution in the ADD model provides a continuum distribution, in the RS model one finds well-separated massive KK resonances. Using the coefficients already obtained in ADD scenario, we first study the invariant mass distribution for DY production at NNLO accuracy in the RS model. Next we study the impact of N$^3$LO SV correction as well as the NNLL resummed effect over the NNLO correction within this model.

The article is organized as follows: in sec.\ (\ref{sec:model}) we briefly describe the RS model and present the interaction lagrangian. In sec.\ (\ref{sec:dyinv}) we   set up the theoretical framework for invariant mass distribution for di-lepton production in RS model followed by the discussion on  SV cross-section in sec.\ (\ref{sec:svxsect}) and threshold resummation in sec.\ (\ref{sec:resum}). In sec.\ (\ref{sec:numerics}), we present the distribution at NNLO and the results for N$^3$LO SV as well as the resummed results matched at NNLO+NNLL accuracy. Finally we conclude in sec.\ (\ref{sec:conclusion}).

\section{Theoretical Framework} \label{sec:theory}
\subsection{The Model}\label{sec:model}
The RS background is a warped metric and can be parametrized \cite{Davoudiasl:1999jd} as 
\begin{align}
ds^{2} = e^{-2\kappa r_{c} |\phi|} \eta_{\mu\nu} dx^{\mu}dx^{\nu} - r_{c}^{2}d\phi^{2} \qquad ,
\end{align}
where $\eta_{\mu\nu}$ is the flat Minkowski metric and $\phi$ is the extra dimension with periodicity $0\leq \phi \leq \pi$ and is compactified on a $S^1/\mathbb{Z}_2$ orbifold with radius $r_c$. The curvature of the $AdS_5$ space-time is denoted as $\kappa$. In the RS model, there are two 3-branes located at two orbifold fixed points on the coordinate of the fifth dimension $\phi = 0$ and $\phi=\pi$ known as the `Planck brane' and `TeV brane' respectively. All the SM particles are confined in the TeV brane and only the gravity is allowed to propagate into the fifth dimension. With this set-up, the hierarchy between the electroweak scale and the Planck scale is understood reasonably if the compactification radius ($r_c$) and the $AdS$ curvature ($\kappa$) satisfy a condition, $\kappa r_c \sim {\cal O}(10)$ \cite{Goldberger:1999un,Goldberger:1999uk}. The higher KK modes therefore will produce observable effects on the LHC processes at the TeV range. These massive KK states ($Y_{\mu\nu}^{(n)}$) interact with the SM fields through stress-energy tensor ($T^{\mu\nu}$) and the interaction Lagrangian is given \cite{Han:1998sg,Giudice:1998ck} as,
\begin{align}
 \mathcal{L}_{RS} = - \frac{1}{\overline{M}_{Pl}} T^{\mu\nu} (x)Y^{(0)}_{\mu\nu}(x) 
 - \frac{\overline{c}_{0}}{m_{0}} T^{\mu\nu} (x)\sum^{\infty}_{n=1} Y^{(n)}_{\mu\nu}(x) \,.
\end{align}
The interaction with zeroth KK mode ($Y_{\mu\nu}^{(0)}$) is suppressed by the reduced Planck mass ($\overline{M}_{Pl}$) and thus can be neglected for practical purposes. The higher KK modes however couple with the strength $\frac{\overline{c}_{0}}{m_{0}}$, where $\overline{c}_{0}=\frac{\kappa}{\overline{M}_{Pl}}$, $m_{0}=\kappa e^{-\kappa r_{c} \pi}$. The masses of the KK modes are given by $M_n = x_n\,\kappa\,e^{-\pi\kappa r_c}$, with $x_n$ being the zeros of the Bessel function $J_1(x)$. The effective graviton propagator can be found after summing over all the massive KK modes except the zeroth one and it takes the  form \cite{Davoudiasl:2000wi,Kumar:2009nn},
\begin{align}
D_{eff}(s_{ij}) &= \sum_{n=1}^{\infty} \frac{1}{s_{ij} - M_n^2 + i \Gamma_n M_n} \nonumber \\
&= \frac{1}{m_{0}^{2}}\sum_{n=1}^{\infty} \frac{\left(x^{2}-x_{n}^{2}\right) 
-i x_{n}\frac{\Gamma_{n}}{m_{0}}}{\left(x^{2}-x_{n}^{2}\right)^{2}
+x_{n}^{2}\left(\frac{\Gamma_{n}}{m_{0}}\right)^{2}} \,,
\end{align}
where $s_{ij}=(p_i+p_j)^2$, $x=\sqrt{s_{ij}}/m_0$ and $\Gamma_n$ denotes the width of the resonance with mass $M_n$ (see \cite{Han:1998sg,KumarRai:2003kk}). In the RS model,  the individual KK resonances are well-separated  and  can be probed for example in the invariant mass distribution of lepton pairs in DY production.

\subsection{Drell-Yan invariant mass distribution}\label{sec:dyinv}
The invariant mass distribution for  DY production at the hadron collider is given by,
\begin{align}\label{eq:dylhc}
2 S~{d \sigma^{} \over d Q^2}\left(\tau,Q^2\right)
&=\sum_{ab={q,\overline q,g}} \int_0^1 dx_1
\int_0^1 dx_2~
\int_0^1 dz \,\, \delta(\tau-z x_1 x_2)
\nonumber\\[2ex] &
\times 
{\cal L}_{ab}(x_1,x_2,\mu_f^2) 
\sum_{I}
\Delta^{I}_{ab}(z,Q^2,\mu_f^2)
\,.
\end{align}
Here $S$ and $\hat{s}$ denote the centre-of-mass energy in the hadronic and partonic frame respectively. The mass factorized partonic coefficient function $\Delta^{I}_{ab}$ is convoluted with the parton luminosity distribution ${\cal L}_{ab}$  consisting of parton distribution functions $f_a^{P_1}(x_1,\mu_f^2)$ and $f_b^{P_2}(x_2,\mu_f^2)$ respectively for two incoming protons. The summation over $I$ takes care of the SM and the RS contributions. The hadronic and partonic threshold variables $\tau$ and $z$ are defined as
\begin{align}
\tau=\frac{Q^2}{S}, \qquad z= \frac{Q^2}{\hat{s}} \,.
\end{align}
They are thus related by $\tau = x_1 x_2 z$. To the all order in strong coupling, the partonic cross-section in the above \eq{eq:dylhc} can be decomposed as the sum of soft-virtual (SV) piece  and regular piece (up to normalization of born contribution),
\begin{align}
\Delta^{I}_{ab}  \equiv  \sum_n \Delta^{I,(n)}_{ab}  = {\cal F}^{(0)}_I \Big(\Delta^{({\rm sv},I)}_{ab} + \Delta^{({\rm reg},I)}_{ab} \Big) \,.
\end{align}
At each order of strong coupling, the SV terms contain all the leading singular terms consisting of `plus-distributions' $\Big[ \frac{\ln^i (1-z)}{(1-z)}\Big]_+$ and delta function $\delta(1-z)$. The regular piece on the other hand is finite in $z$.

The pre-factor ${\cal F}^{(0)}$ takes the following form for neutral vector bosons in SM  and spin-2 (RS) boson respectively,
\begin{align}
{\cal F}_{\rm SM}^{(0)} &=
{4 \alpha^2 \over 3 Q^2} \Bigg[Q_q^2 
- {2 Q^2 (Q^2-M_Z^2) \over  \left((Q^2-M_Z^2)^2
+ M_Z^2 \Gamma_Z^2\right) c_w^2 s_w^2} Q_q g_e^V g_q^V  \nn\\
&+ {Q^4 \over  \left((Q^2-M_Z^2)^2+M_Z^2 \Gamma_Z^2\right) c_w^4 s_w^4}\Big((g_e^V)^2
+ (g_e^A)^2\Big)\Big((g_q^V)^2+(g_q^A)^2\Big) \Bigg]\,, \nn\\
{\cal F}^{(0)}_{\rm RS} &= \frac{2Q^2}{\Lambda_\pi^2}\,,
\end{align}
where $\alpha$ is the fine structure constant, $Q$ is the invariant mass of the lepton pair, $M_Z$ and $\Gamma_Z$ are the mass and the decay width of the $Z$-boson,  $c_w,s_w$ are sine and cosine of Weinberg angle respectively. The vector and axial-vector part of the weak boson coupling is given as,
\begin{align}
g_a^A = -\frac{1}{2} T_a^3 \,, \qquad g_a^V = \frac{1}{2} T_a^3  - s_w^2 Q_a \,,
\end{align}
$Q_a$ being electric charge and $T_a^3$ is the weak isospin of the electron or quarks. Note that the SM contribution consists of contribution from $\gamma$ and $Z$ as well as their interference. For the invariant mass distribution, however the spin-2 production is decoupled from the SM one \cite{Mathews:2004xp} and thus there is no interference of them. 

The complete SM contribution to DY invariant mass distribution is known up to second order in the strong coupling \cite{Hamberg:1990np,Altarelli:1978id,Matsuura:1987wt,Matsuura:1988sm}. Up to two loops the contribution from RS spin-2 can be written as,
\begin{align}
	2 S{\frac{d \sigma^{}_{RS}}{dQ^2}}(\tau,Q^2)&=
	\sum_{q,\bar q,g}{\cal F}^{(0)}_{RS} \int_0^1 {d x_1 } \int_0^1 
{dx_2} \int_0^1 dz~ \delta(\tau-z x_1 x_2) 
\nonumber\\
\times
&\Bigg[ 
{\cal L}_{q{\bar q}} 
             \sum\limits_{n=0}^{2} a_{s}^{n} \Delta^{RS, (n)}_{q{\bar q}} 
+
{\cal L}_{g g} \sum\limits_{n=0}^{2} a_{s}^{n} \Delta^{RS, (n)}_{gg} 

\nn\\&
           + \Big( {\cal L}_{gq} + {\cal L}_{qg}  \Big)  \sum\limits_{n=1}^{2} a_{s}^{n} \Delta^{RS, (n)}_{gq}
\nonumber\\&+
{\cal L}_{q q} \sum\limits_{n=2}^{2}
             a_{s}^{n} \Delta^{RS, (n)}_{qq}  
+ {\cal L}_{q_{1} q_{2}}  \sum\limits_{n=2}^{2}
             a_{s}^{n} \Delta^{RS, (n)}_{q_{1}q_{2}}  
\Bigg]\,,
\end{align}
with 
\begin{align}
\label{eq:32}
\L_{q \bar q}(x_1,x_2,\mu_f^2)&=
f_q^{P_1}(x_1,\mu_f^2) 
f_{\bar q}^{P_2}(x_2,\mu_f^2)
+f_{\bar q}^{P_1}(x_1,\mu_f^2)~ 
f_q^{P_2}(x_2,\mu_f^2)\,,
\nonumber\\
\L_{q q}(x_1,x_2,\mu_f^2)&=
f_q^{P_1}(x_1,\mu_f^2) 
f_{q}^{P_2}(x_2,\mu_f^2)
+f_{\bar q}^{P_1}(x_1,\mu_f^2)~ 
f_{\bar q}^{P_2}(x_2,\mu_f^2)\,,
\nonumber\\
\L_{q_1 q_2}(x_1,x_2,\mu_f^2)&=
f_{q_1}^{P_1}(x_1,\mu_f^2) 
\Big( f_{q_2}^{P_2}(x_2,\mu_f^2) + f_{\bar q_2}^{P_2}(x_2,\mu_f^2) \Big)
\nonumber\\&+f_{\bar q_1}^{P_1}(x_1,\mu_f^2)~ 
\Big( f_{q_2}^{P_2}(x_2,\mu_f^2) + f_{\bar q_2}^{P_2}(x_2,\mu_f^2) \Big)\,,
\nonumber\\
\L_{g q}(x_1,x_2,\mu_f^2)&=
f_g^{P_1}(x_1,\mu_f^2) 
\Big(f_q^{P_2}(x_2,\mu_f^2)
+f_{\bar q}^{P_2}(x_2,\mu_f^2)\Big)\,,
\nonumber\\
\L_{q g}(x_1,x_2,\mu_f^2)&=
\L_{g q}(x_2,x_1,\mu_f^2)\,,
\nonumber\\
\L_{g g}(x_1,x_2,\mu_f^2)&=
f_g^{P_1}(x_1,\mu_f^2)~ 
f_g^{P_2}(x_2,\mu_f^2)\,.
\end{align}

Notice that computation of the partonic coefficients  at the second order requires evaluation of matrix element as well as the proper phase space for the di-lepton pair. Using the method of reverse unitarity \cite{Anastasiou:2002yz}, where the phase space integrals were converted to loop integrals, the later has been performed very recently in the case of generic spin-2 production \cite{Ahmed:2016qhu}. The advantage is that, one then can re-use all the techniques developed for the multi-loop computation. The analytical result obtained in this way is useful for any spin-2 production with universal coupling to the SM. We use these coefficients to predict complete NNLO cross-section for the RS model. 

\subsection{Soft-Virtual cross-section}\label{sec:svxsect}

It is important to consider corrections beyond NNLO in order to obtain perturbative stability. The first step as discussed earlier is to calculate the third order soft-virtual cross-section. This can be achieved by calculating the spin-2 form factor at three loops \cite{Ahmed:2015qia} as well as the soft function at the same order. The soft function being maximally non-abelian up to three loops, can be extracted from the known Higgs \cite{Anastasiou:2014vaa,Li:2014afw} and DY \cite{Ahmed:2014cla,Catani:2014uta} results. Using these informations, the third order coefficients for the SV corrections have been obtained \cite{Das:2019bxi} for generic spin-2 coupling and have been applied to ADD model to predict the DY distribution to N$^3$LO.  The analytical coefficients can be also used to predict the SV cross-section in the RS graviton production at the same perturbative order. The SV cross-section at each order consists of plus-distributions and delta function which are most singular terms in the partonic coefficient function. One can write the SV coefficient in terms of perturbative expansion in strong coupling,
\begin{align}
\Delta_{ab}^{{(\rm sv},I)} &= \sum_{n=0}^{\infty} a_s^n \Delta_{ab}^{(n),I}\,.
\end{align}
The SV coefficients arise only in flavor diagonal channels \ie either in $q\bar{q}$ or $gg$ born process. At any order $n$ of the strong coupling, the SV coefficients have the following structure in terms of plus distributions and delta function,
\begin{align}
\Delta_{ab}^{(n),I} &= c_{n,\delta}^I ~\delta(1-z) + \sum_{i=0}^{2n-1} c_{n,i}^I ~ \bigg[\frac{\ln^i(1-z)}{(1-z)} \bigg]_+  \,.
\end{align}
In the SM, only $q\bar{q}$ contributes wheres for the RS scenario both $q\bar{q}$ and $gg$ channels contribute to the SV coefficient. Here we present only the leading order term (\ie $n=0$) in this series to follow the overall normalization to the coefficient,
\begin{align}
\Delta_{\rm q\bar{q}}^{(0),SM} &= \frac{2\pi}{n_c} \delta(1-z)\,, \nn\\
\Delta_{\rm q\bar{q}}^{(0),RS}&=  \frac{\pi}{8 n_c} \delta(1-z)\,, \nn\\
\Delta_{\rm gg}^{(0),RS} &= \frac{\pi}{2 (n_c^2-1)} \delta(1-z)
\end{align}
Up to two loops, these are already known for quite some time and can be found in \cite{deFlorian:2013wpa}. Recently we calculated the three loop pieces \cite{Das:2019bxi} for generic spin-2 couplings. In this article we use these three-loop coefficients for the third order phenomenological prediction in the RS model.

\subsection{Threshold resummation}\label{sec:resum}

The NNLO cross-section can be improved with the contribution from threshold logarithms at all orders. In particular when partonic $z\to 1$ the contribution from these singular terms becomes large and unreliable and thus needs to be resummed to all orders. The resummation is usually performed in conjugate space where all the convolutions become simple product and therefore easy to calculate. We follow the standard approach where we evaluate the resummed coefficients in the Mellin-$N$ space. The threshold limit translates there into $\Nb \to\infty$ with $\Nb = N\exp(\gamma_E^{})$, $N$ being Mellin conjugate to $z$ and $\gamma_E^{}$ is the Euler-Mascheroni constant. Up to the born factor, the partonic resummed cross-section can be organized as \cite{Catani:1996yz,Catani:1989ne,Moch:2005ba} ,
\begin{align}\label{eq:resum-parton}
(d\hat{\sigma}_N/dQ)/(d\hat{\sigma}_{\rm LO}/dQ) = g_0^I \exp \Big( G_{\Nb}^I \Big) \,.
\end{align}
The normalization $(d\hat{\sigma}_{\rm LO}/dQ)$ is given as,
\begin{align}
(d\hat{\sigma}_{\rm LO}/dQ) &= {\cal F}^{(0)}_{\rm SM} \frac{Q}{S} \bigg\{ \frac{2\pi}{n_c}\bigg\} \, ~ \qquad\qquad\qquad~ \text{for SM,} \nn\\
 &= {\cal F}^{(0)}_{\rm RS} \frac{Q}{S} \bigg\{ \frac{\pi}{8 n_c}, \frac{\pi}{2(n_c^2-1)}\bigg\} \qquad \text{for} ~~ \{q\bar{q},gg\}~ \text{in RS.}
\end{align}
The exponent can be found through the following representation in Mellin space in terms of universal constants the cusp anomalous dimensions $A^I$  \cite{Moch:2004pa, Lee:2016ixa, Moch:2017uml, Grozin:2018vdn, Henn:2019rmi, Bruser:2019auj, Davies:2016jie, Lee:2017mip, Gracey:1994nn, Beneke:1995pq, Moch:2017uml, Moch:2018wjh, Henn:2019rmi, Lee:2019zop, Henn:2019swt}  and constants $D^I$ \cite{Catani:2003zt,Moch:2005ba,Das:2019bxi},
\begin{align}
G_{\Nb}^I = \int_0^1 dz \frac{z^{N-1} - 1}{1 - z} \bigg[ \int_{\muf^2}^{Q^2(1-z)^2} \frac{d\mu^2}{\mu^2} 2~A^I(a_s(\mu^2)) + D^I(a_s(Q^2(1-z)^2))\bigg] \,.
\end{align}
Performing the integration, one can organize it as a resummed series realizing $\ln \Nb \sim 1/a_s$,
\begin{align}\label{eq:resum-exponent}
G_{\Nb}^I = \ln \Nb ~g_1^I(\omg) + g_2^I(\omg) + a_s ~g_3^I(\omg) + a_s^2~ g_4^I(\omg) + \cdots \,,
\end{align}
where $\omg = 2 \beta_0 a_s \ln \Nb$. The expressions for $g^I_i$  required up to N$^3$LL resummation can be found in \cite{H.:2020ecd,Das:2019btv,H:2019dcl}. The process dependent coefficient $g_0^I$ can be extracted from the SV results in Mellin-$N$ space and can be written as a perturbative series in strong coupling as,
\begin{align}\label{eq:g0b}
 g_0^I = 1 + \sum_{n=1}^{\infty} a_s^n g_{0n}^I \,.
\end{align}
We have extracted these coefficients up to the third order from the known SV coefficients at the same order for a generic spin-2 coupling, which can be found in \cite{Das:2019bxi}. The first term in \eq{eq:resum-exponent} along with first term in \eq{eq:g0b} define the leading logarithmic (LL) accuracy. Similarly the first two terms in \eq{eq:resum-exponent} along with the terms up to ${\cal O}(a_s)$ in \eq{eq:g0b} define next-to-leading logarithm (NLL) order and so on. Note that the expansion in \eq{eq:resum-exponent} is different from \cite{Moch:2005ba,Catani:2003zt}, where one organizes the series in terms of $N$ instead of $\Nb$.  Physically both are equivalent in the large-$N$ limit, however numerically it has been seen that the $\Nb$ exponentiation shows better perturbative convergence for DIS \cite{Das:2019btv} as well as in DY \cite{H.:2020ecd}. 

The partonic resummed cross-section has to be Mellin-inverted with suitable $N$-space PDF and finally has to be matched with the known fixed order results. The general expression for the matched cross-section can be written as below: 
\begin{align}\label{eq:matched}
\bigg[\frac{d\sigma}{dQ}\bigg]_{N^nLO+N^nLL} &= \bigg[\frac{d\sigma}{dQ}\bigg]_{N^nLO} +
\sum_{ab\in\{q,\bar{q},g\}}
 \frac{d\hat{\sigma}_{LO}}{dQ}  \int_{c-i\infty}^{c+i\infty} \frac{dN}{2\pi i} (\tau)^{-N} 
 \nn\\
& \delta_{ab}f_{a,N}(\muf^2) f_{b,N}(\muf^2) 
\times \bigg( \bigg[ \frac{d\hat{\sigma}_N}{dQ} \bigg]_{N^nLL} - \bigg[ \frac{d\hat{\sigma}_N}{dQ} \bigg]_{tr}     \bigg) \,.
\end{align}
The first term in the above \eq{eq:matched} represents the fixed order known to N$^n$LO. The last term in the bracket represents the resummed result truncated to the fixed order accuracy \ie to N$^n$LO to remove all double counting of the singular terms that are already present in  fixed order. $f_{i,N}(\muf^2)$ are the Mellin space PDFs which can be evolved using publicly available code  QCD-P{\sc egasus} \cite{Vogt:2004ns}, however we followed the prescription provided in \cite{Catani:1996yz} relating $N$-space PDFs to the derivative of $x$-space PDF for simplicity and flexibility to using {\tt lhapdf} \cite{Buckley:2014ana} libraries. To avoid the Landau pole problem in the Mellin-inversion integration, we have followed the minimal prescription \cite{Catani:1996yz} and chosen the contour accordingly.  All the necessary analytical ingredients are now available to perform the numerical study which we report in the next section.
In our work, all the algebraic computations have been done with the latest version of the symbolic manipulation system {\sc Form}~\cite{Vermaseren:2000nd,Ruijl:2017dtg}. 

\section{Numerical Results}\label{sec:numerics}

In this section we present our numerical results for the di-lepton production cross section in the RS model at LHC.
The LO, NLO and NNLO parton level cross sections are convoluted with the respective order by order parton distribution 
functions (PDF) taken from {\tt lhapdf} \cite{Buckley:2014ana}. The corresponding strong coupling constant 
$a_s(\mu_r^2) = \alpha_s(\mu_r^2)/(4\pi)$ is also provided by {\tt lhapdf}. The fine structure constant 
and the weak mixing angles are chosen to be  
$\alpha_{\rm em} = 1/128$ and $\sin^2\theta_w = 0.227$ respectively. Here the results are presented for 
$n_f = 5$ flavors in the massless limit of quarks.  The default choice for the center of mass energy of protons is 13 TeV 
and the choice for the PDF set is MMHT2014 \cite{Harland-Lang:2014zoa}. Except for the scale variations, we have used the 
factorization ($\mu_f$) and  
renormalisation ($\mu_r$) scales to be the invariant mass of the di-lepton, i.e. $\mu_f = \mu_r = Q$. We also note that 
there have been several experimental searches at the LHC for warped extra dimensions in the past, yielding stringent bounds on 
the RS model parameters, the mass of the first resonance mode ($M_1$) and the coupling strength ($\bar{c}_{0}$)\cite{Khachatryan:2016zqb,Aaboud:2017yyg}.
Such analyses have already used the K-factors that have been computed in the extra dimension models. 
Here in this work, for our phenomenological study to assess the impact of QCD corrections,  we choose $M_1 = 1.5$ TeV and 
$\bar{c}_{0} = 0.05$. The computational details of the QCD corrections presented here are model independent,  
and a numerical estimate of the theory predictions for  any other choice of the model parameters is straight-forward. 
For completeness, we also study the dependence of the invariant mass distributions on the model parameters considering 
the recent bounds on $M_1$ for different $\bar{c}_{0}$ values.


\subsection{Fixed order corrections}\label{sec:fo}

\begin{figure}[ht!]
  \centering

                   {\includegraphics[trim=0 0 0 0,clip,width=0.55\textwidth,height=0.48\textwidth]
                   {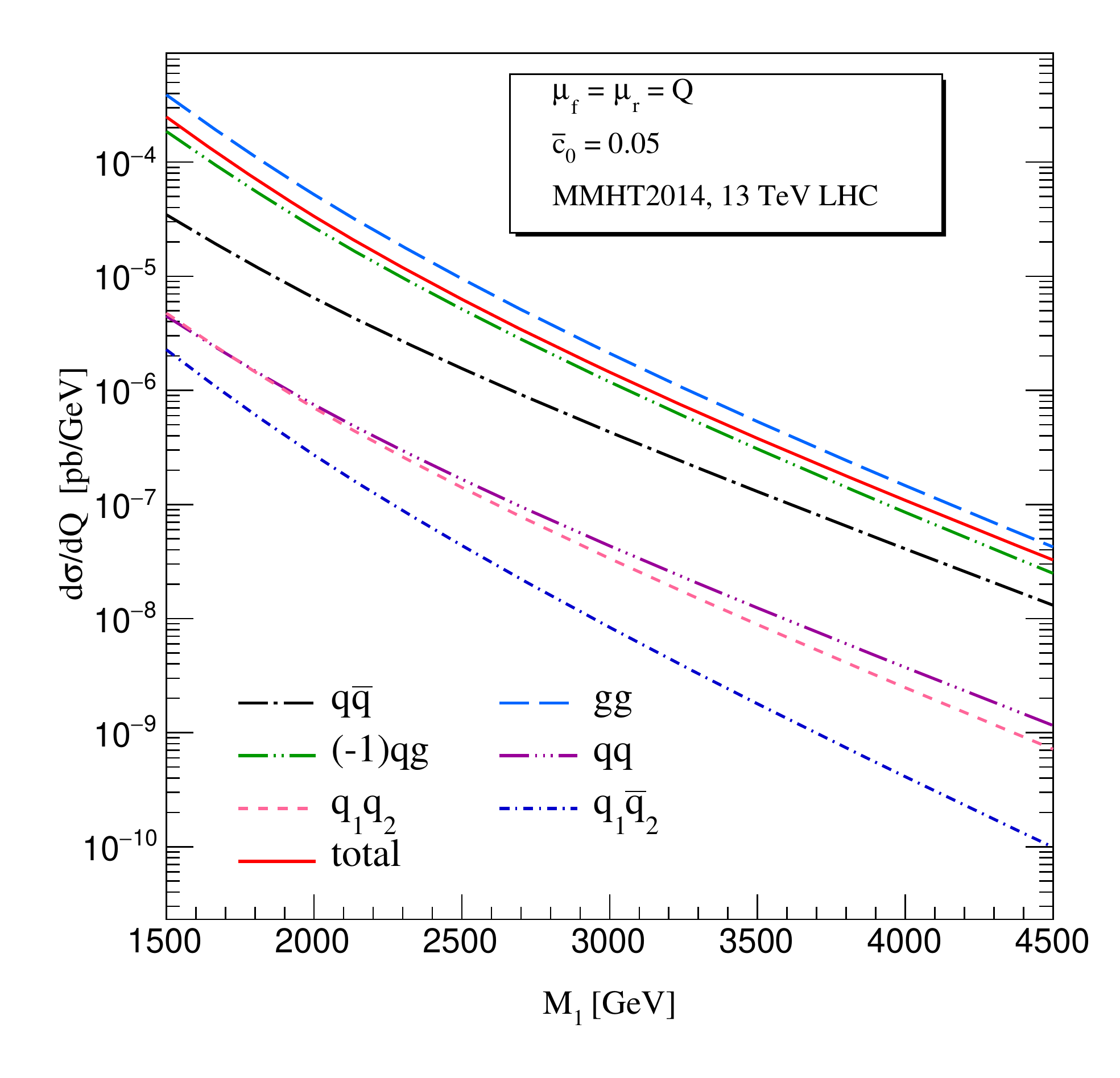}}
	\caption{Different subprocess contributions for the RS model at NNLO QCD right at the resonance for different $M_1$
		 values keeping $\bar{c}_{0}$ fixed at $0.05$.}
\label{subprocess_contribution}
\end{figure}
First we present in \fig{subprocess_contribution} the contribution from different subprocesses for the pure RS graviton (GR) at NNLO level right at the resonance region by  varying the first resonant mass $M_1$ and keeping $\bar{c}_{0} = 0.05$.  At this order in QCD there are six different subprocesses that contribute for GR case, \viz $q\bar{q}$, $gg$, $qg$, $qq$, $q_1q_2$ and $q_1\bar{q}_2$.  Here, the dominant contribution comes from the $gg$-subprocess and it remains dominant for resonance values as large as $4.5$ TeV. The next dominant contribution comes from $qg$-subprocess but it is negative for this entire mass ranges. This is followed by quark initiated processes with $q\bar{q}$ being the largest in this category. For a typical choice of first resonance $M_1 = 2500$, we find that the total cross-section is $0.63 \times 10^{-5}$ pb in which the dominant $gg$ subprocess overshoots the value by $151\%$. The $q\bar{q}$, $qq$, $q_1q_2$ and $q_1\bar{q}_2$ channels contribute in addition $24.7\%$, $2.7\%$, $2.2\%$, $0.7\%$ respectively of the total cross-section. As stated earlier only the $qg$ channel contributes negatively of about $-82\%$ of the total 
cross-section.

\begin{figure}[ht!]
  \centering

                   {\includegraphics[trim=0 0 0 0,clip,width=0.48\textwidth,height=0.40\textwidth]
                   {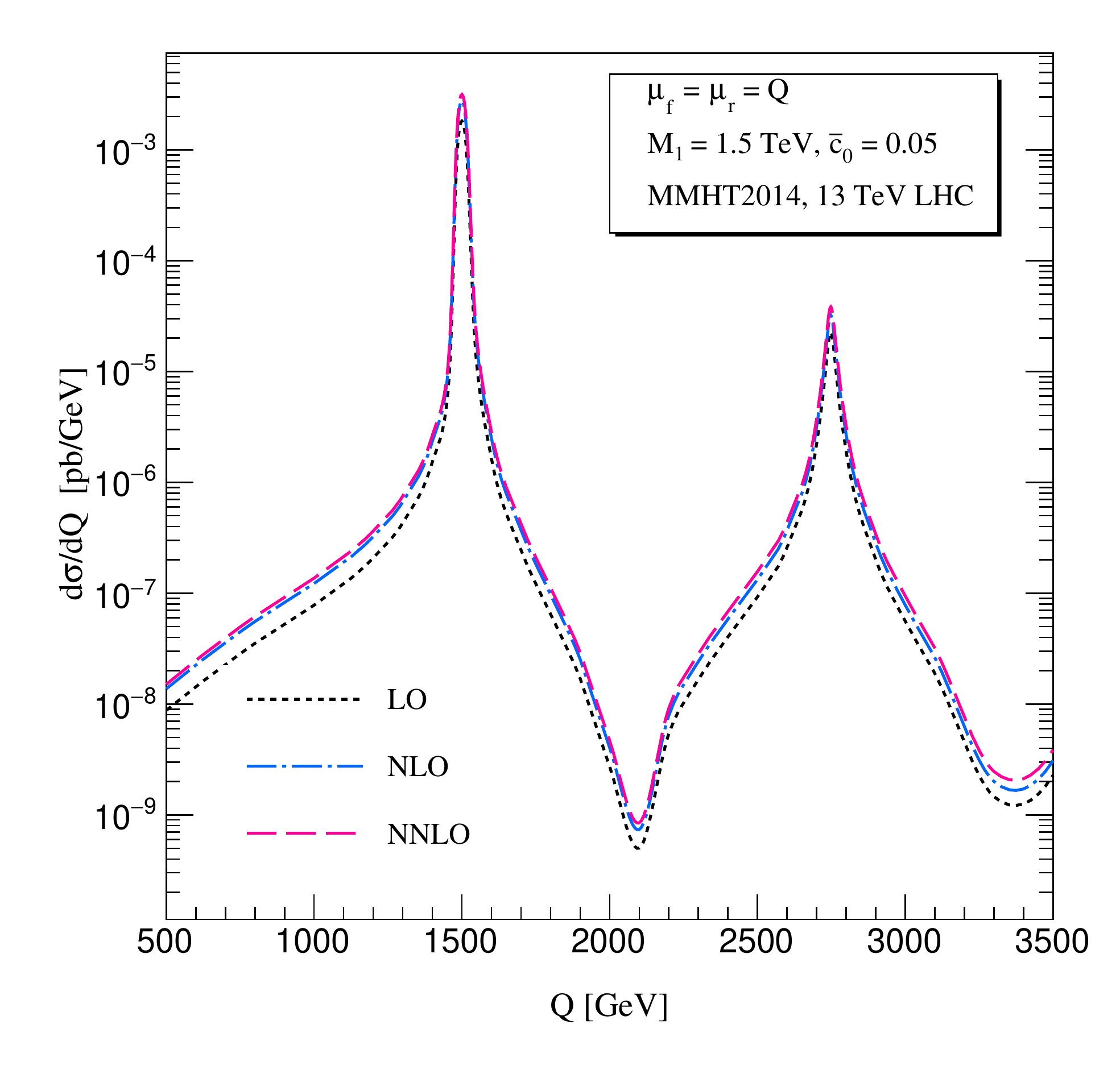}}
  \hskip0.05cm
                   {\includegraphics[trim=0 0 0 0,clip,width=0.48\textwidth,height=0.40\textwidth]
                   {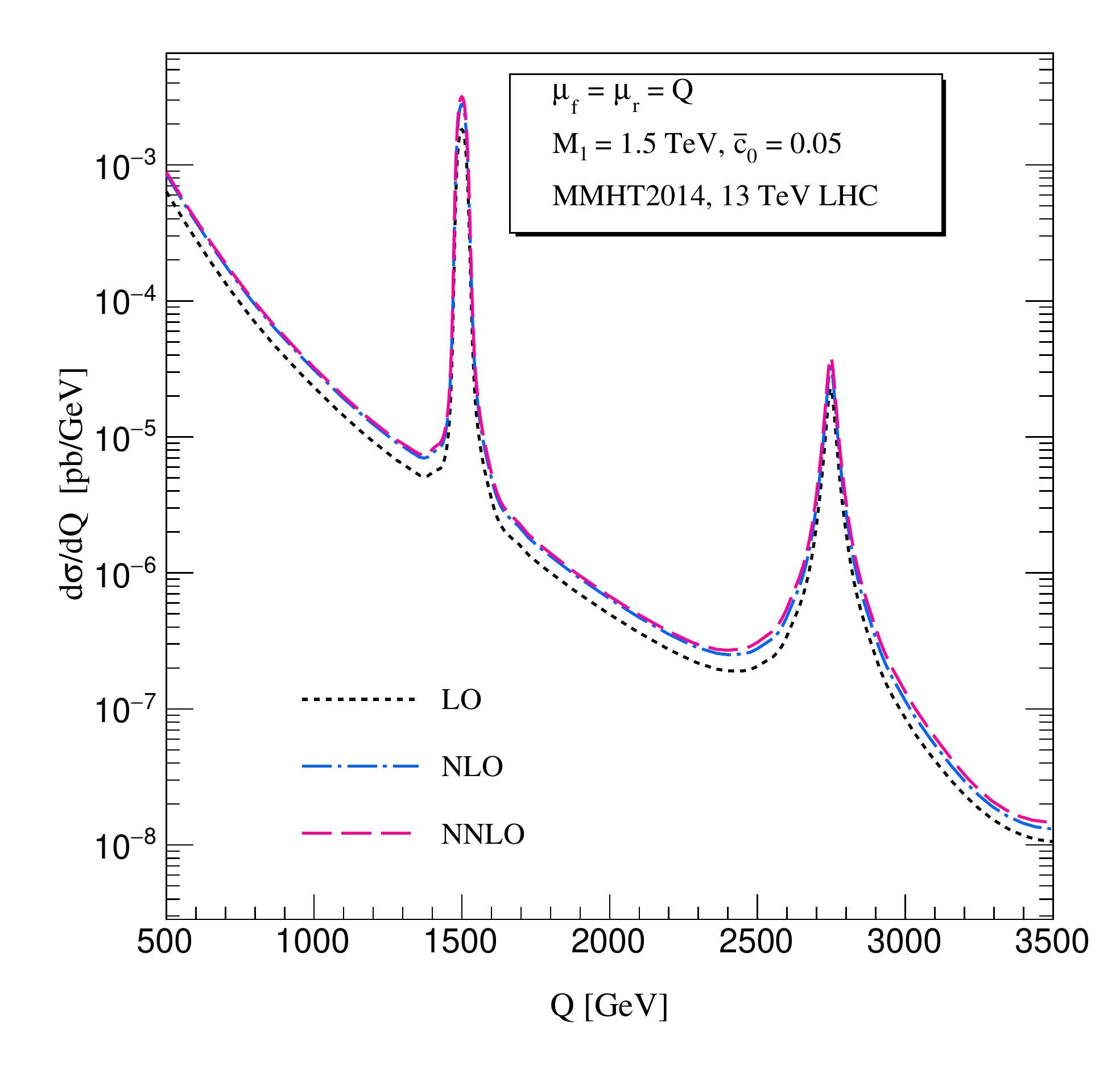}}

	\caption{Di-lepton invariant mass distribution up to NNLO QCD for pure RS model (left panel) and for the signal 
(right panel).}
\label{nnlo_inv_distribution}
\end{figure}
Next, we present in \fig{nnlo_inv_distribution} the di-lepton invariant mass distribution ($d\sigma/dQ$) 
as a function of the invariant mass of the di-lepton $Q$ for GR and for the signal (SM+GR). 
The width of the resonance depends on $\bar{c}_{0}$ and near the resonance region the signal receives most of the contribution 
from the pure RS graviton. Far away from this resonance region, the RS contribution is found to be comparable to that of the SM 
background for $Q > 3500$ GeV.

\begin{figure}[h!]
  \centering
                   
                   {\includegraphics[trim=0 0 0 0,clip,width=0.48\textwidth,height=0.40\textwidth]
                   {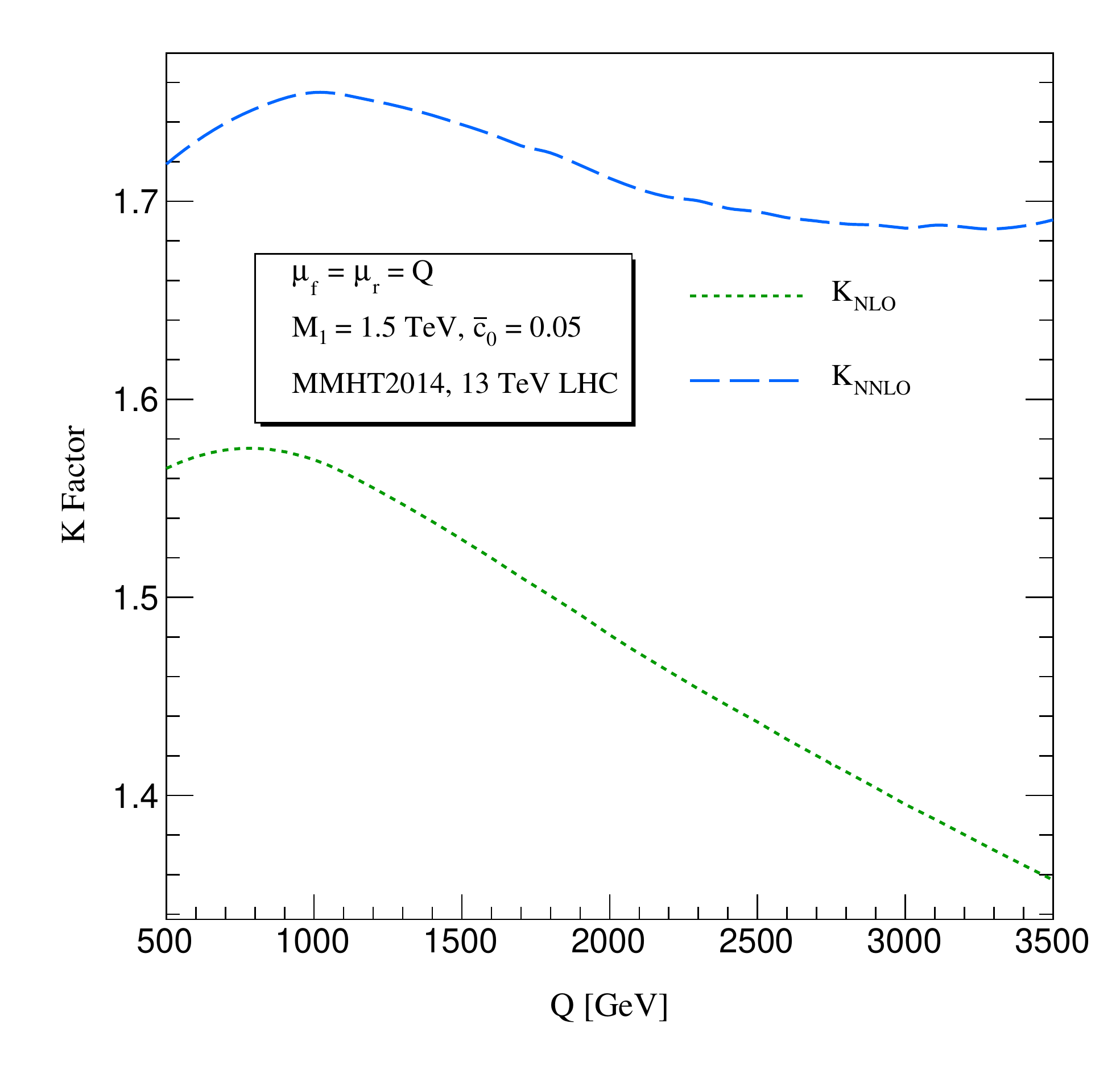}}
  \hskip0.05cm
                   {\includegraphics[trim=0 0 0 0,clip,width=0.48\textwidth,height=0.40\textwidth]
                   {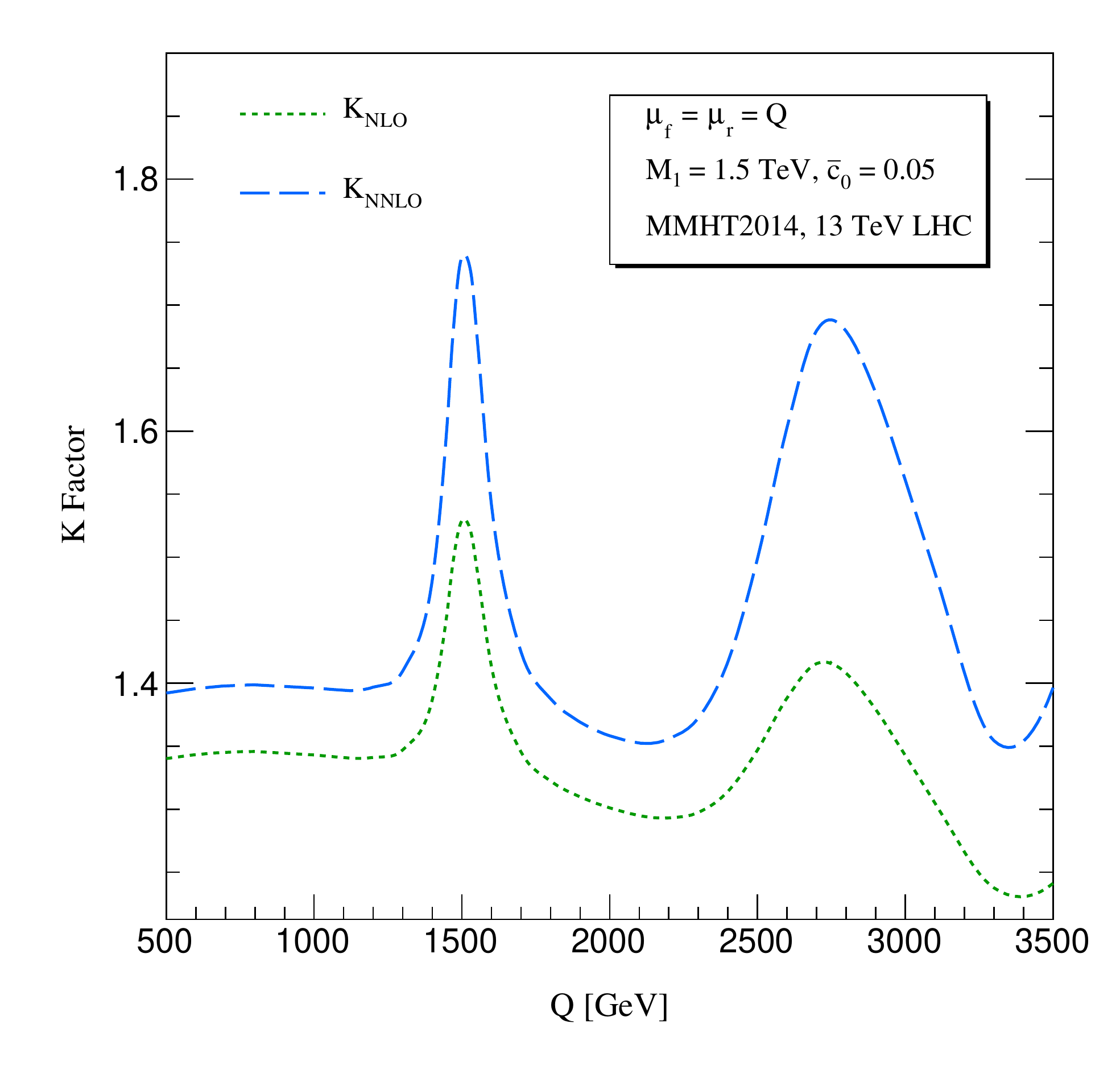}}

\caption{The K-factors up to NNLO in QCD for RS model (left panel) and for the signal (right panel).}
\label{nnlo_k_distribution}
\end{figure}

\begin{align}\label{eqk}
\text{K}_\text{NLO} = \frac{ d\sigma^{\rm NLO}/dQ} {d\sigma^{\rm LO}/dQ } \,,~~
\text{K}_\text{NNLO} = \frac{ d\sigma^{\rm NNLO}/dQ} {d\sigma^{\rm LO}/dQ } \,.
\end{align}
In \fig{nnlo_k_distribution}  we present the K-factor, defined in \eq{eqk}, for both GR and the signal cases. 
In an earlier work we had presented third order SV and resummed results  for di-lepton production at LHC in the ADD model \cite{Das:2019bxi}. We note that it is the same virtual graviton exchange process that contributes both in ADD and RS model. The leading order processes are similar and the QCD corrections are model independent. However, the difference between these two models arise because of the difference in the summation over the tower of KK gravitons and  also in the overall wrapped factor. Consequently the relative weight of the contribution from the gravitons  in these two models will be different for different invariant mass region. This results in different mass-dependent K-factors in the ADD and RS model. The NLO corrections for pure RS case at $Q=1000$ GeV are found to contribute by about 57\% of LO, while NNLO corrections add an additional $18\%$ of LO to the total invariant mass distribution. In \tab{table1} we present the signal K-factors up to NNLO QCD for different $Q$ values. For signal case, the NLO corrections at $Q=1000$ GeV contribute by about 34\% of LO and NNLO corrections add an additional $6\%$ of LO to the total invariant mass distribution. However, right at the resonance region, these NNLO corrections are found to enhance the production cross section by an additional 20\% of LO results. This shows that NNLO corrections are indeed essential for this process in order to make any reliable predictions.
\input{table1.tex}
%
\begin{figure}[h!]
  \centering

                   {\includegraphics[trim=0 0 0 0,clip,width=0.48\textwidth,height=0.40\textwidth]
                   {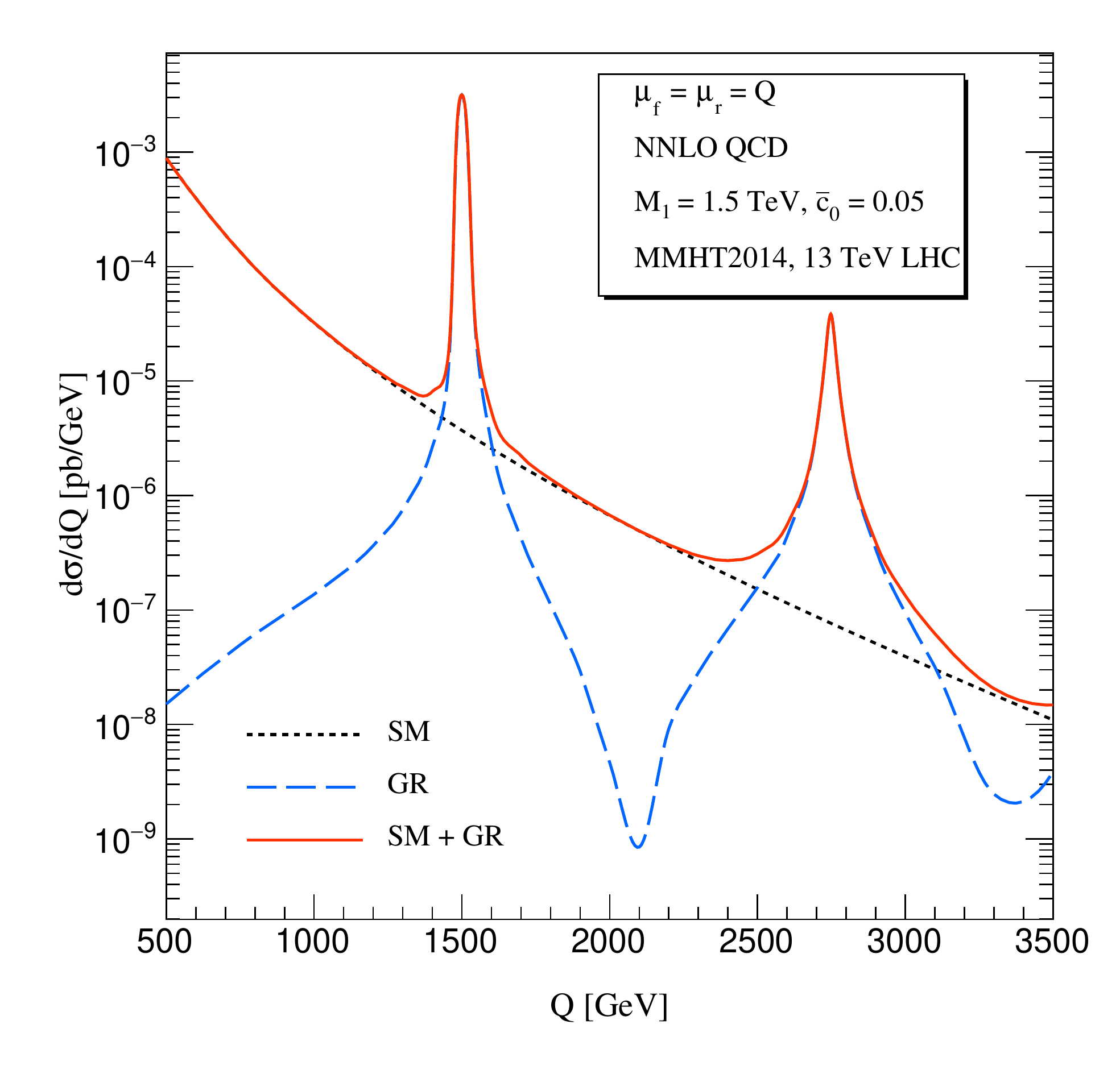}}
  \hskip0.05cm
                   {\includegraphics[trim=0 0 0 0,clip,width=0.48\textwidth,height=0.40\textwidth]
                   {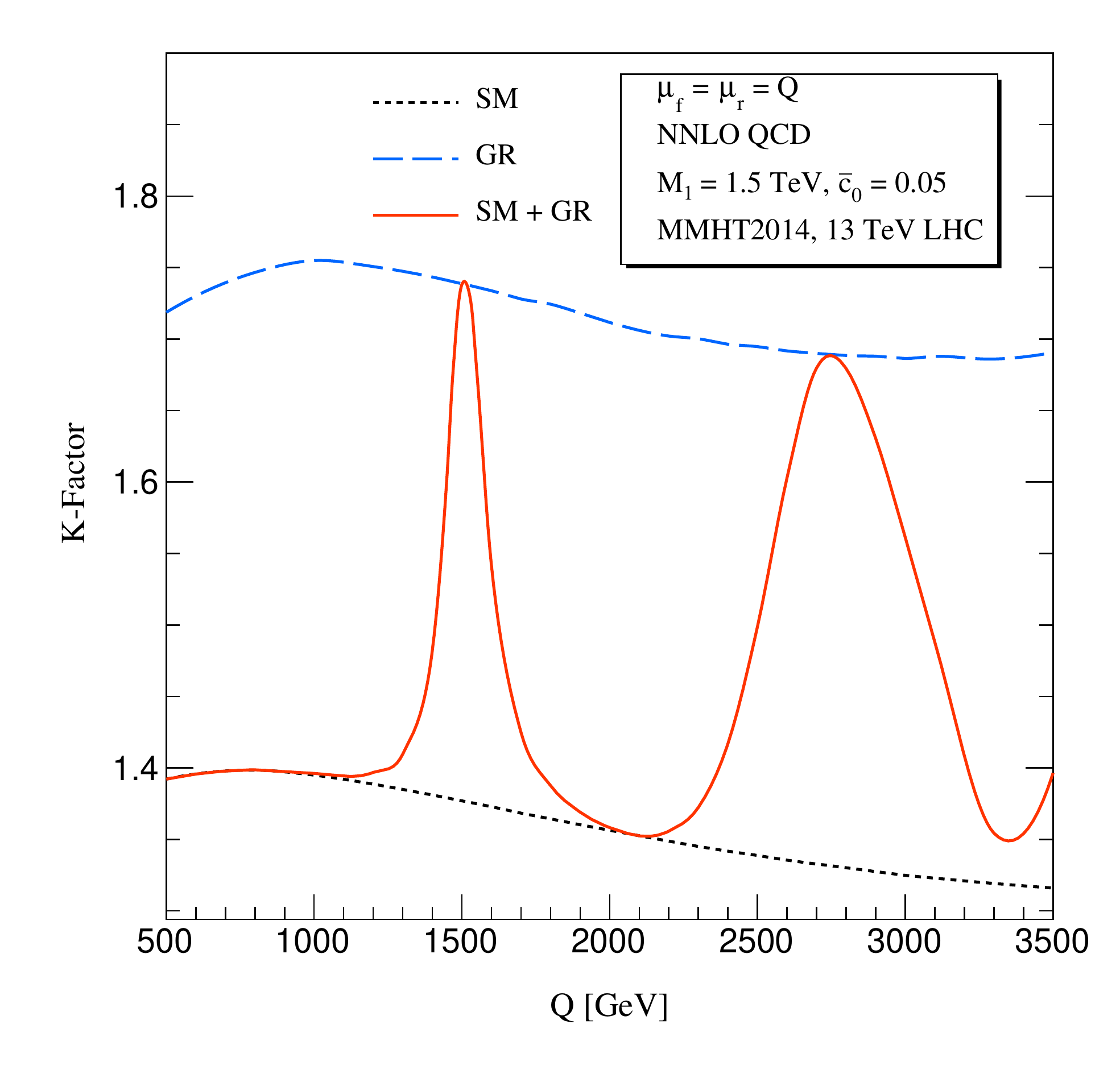}}

\caption{Invariant mass distribution of di-lepton for the SM, RS model and for the signal
        (left) and their corresponding K-factors (right).}
\label{nnlo_model_compare}
\end{figure}

In \fig{nnlo_model_compare}  we present the di-lepton invariant mass distribution for SM, GR and signal cases. The behavior of the signal K-factor is governed by the respective coupling constants in SM and RS as well as the parton fluxes.  As discussed earlier, in the RS case gravity contribution is significant near resonance region and therefore the whole signal K-factor is controlled by RS. In the off resonance region at high $Q$, both RS and SM contributions are comparable and hence the signal K-factor receives contributions from both RS and SM. Hence the behavior of the mass dependent K-factor for the signal in the RS model is very distinct from that in the ADD model \cite{Das:2019bxi}.


\begin{figure}[h!]
  \centering

                   {\includegraphics[trim=0 0 0 0,clip,width=0.48\textwidth,height=0.40\textwidth]
                   {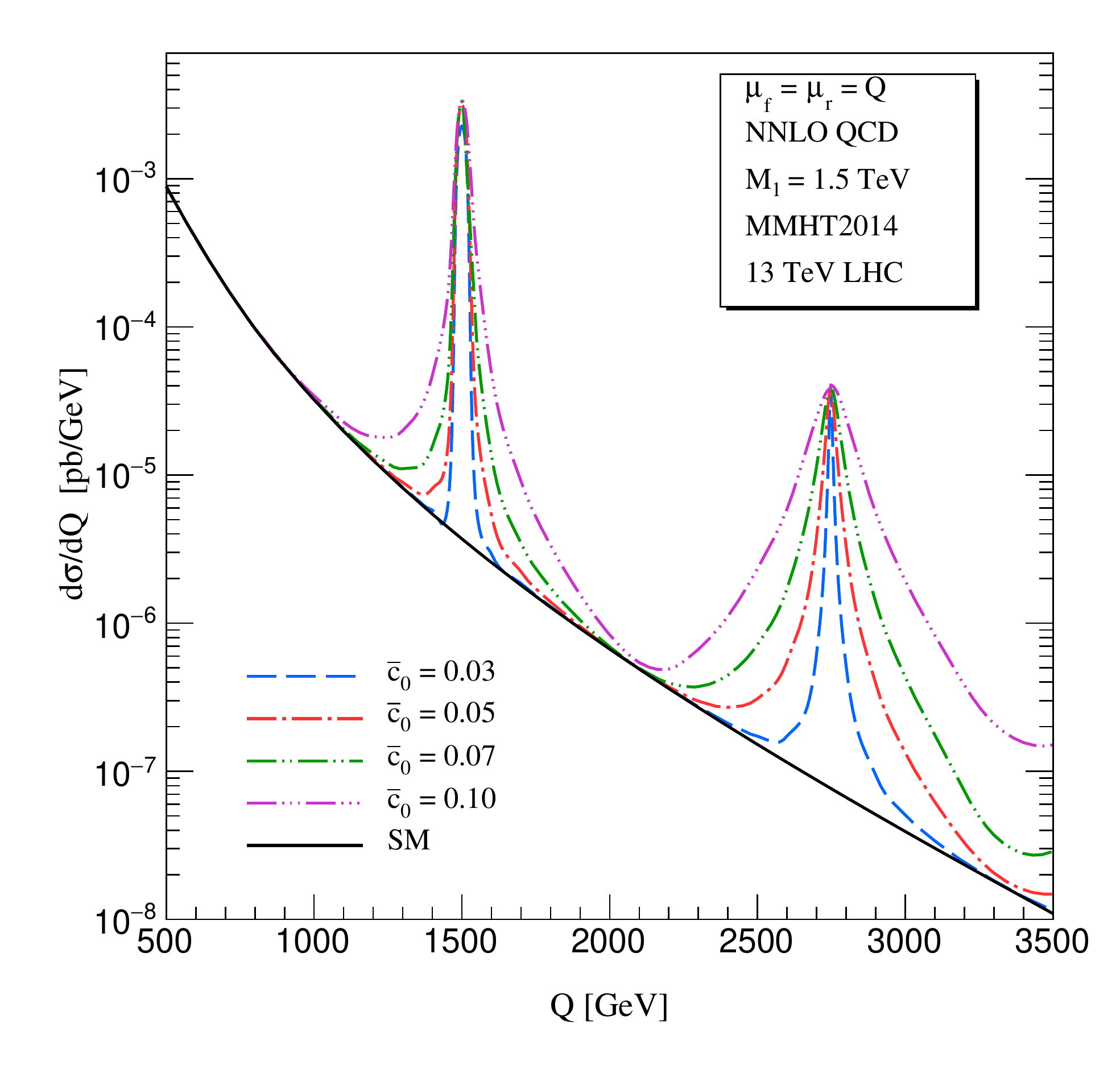}}
  \hskip0.05cm
                   {\includegraphics[trim=0 0 0 0,clip,width=0.48\textwidth,height=0.40\textwidth]
                   {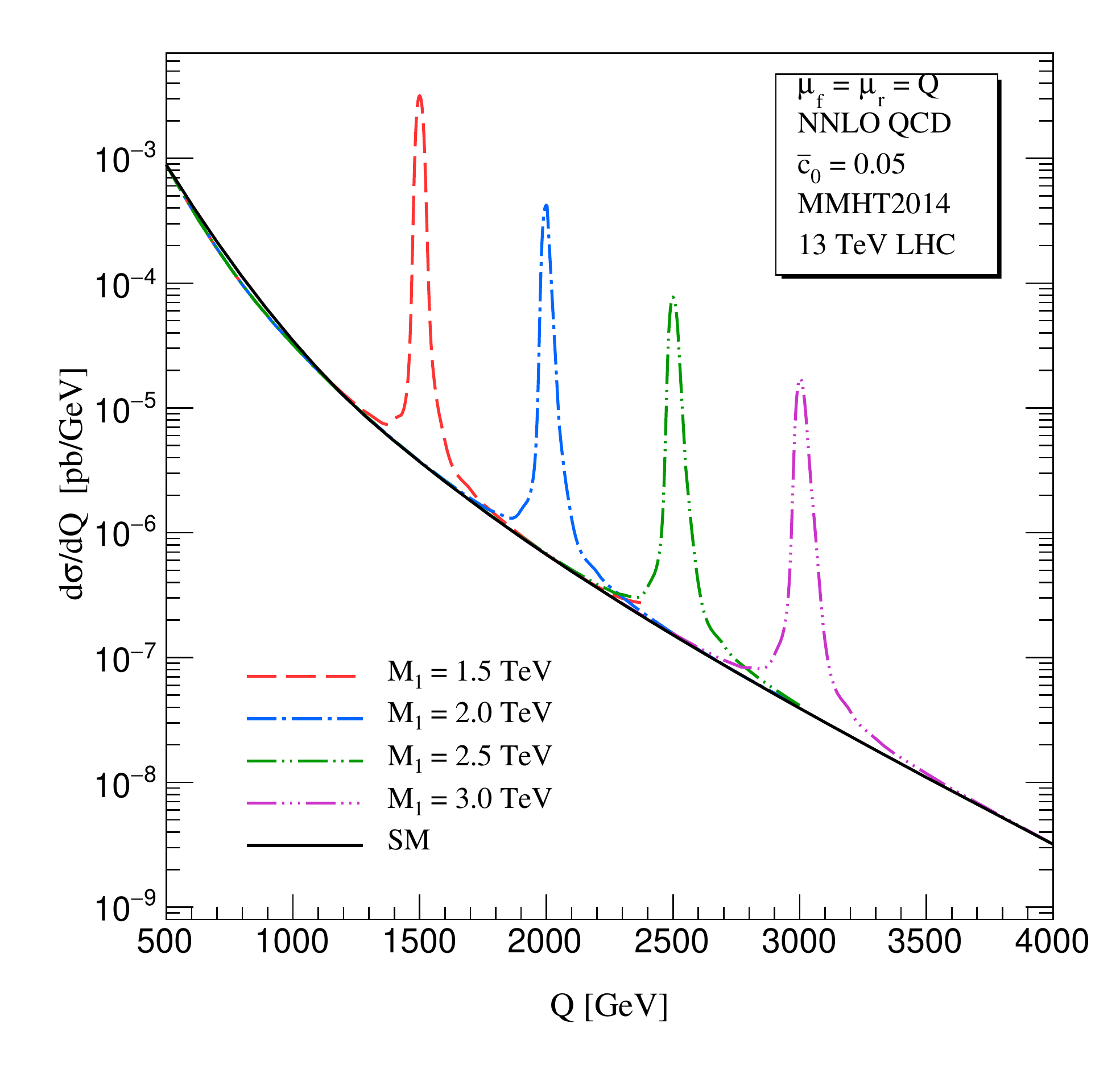}}

\caption{Dependence of the di-lepton invariant mass distribution for the signal on the RS model parameters
         $\bar{c}_0$ (left) and the first resonance mass $M_1$ (right).} 
\label{nnlo_model_parameters}
\end{figure}
%
\input{table2.tex}
We also study the dependence of our results on the RS model parameters, $M_1$ and $\bar{c}_0$. In \fig{nnlo_model_parameters} we present the di-lepton invariant mass distribution by varying $\bar{c}_0$ from $0.03$ to $0.1$ keeping $M_1$ fixed at $1.5$ TeV in the left panel. We also present the results in the right panel by varying $M_1$ from $1.5$ TeV to $3.0$ TeV for a fixed $\bar{c}_0$ ($0.05$). The width of the resonance depends on $\bar{c}_0$, however, right at the resonance this dependence of the production cross section on this coupling $\bar{c}_0$ cancels and the height of peak for any given $M_1$ will be independent of $\bar{c}_0$. Consequently, the respective NNLO K-factors near the resonance region depend on $\bar{c}_0$ but right at the resonance they do not. These signal K-factors are presented in \tab{table2} right at the resonance region for different $M_1$ values. The NNLO corrections increases the K-factors substantially compared to NLO implying the importance of the higher order correction for this process.

\begin{figure}[h!]
  \centering

                   {\includegraphics[trim=0 0 0 0,clip,width=0.60\textwidth,height=0.50\textwidth]
                   {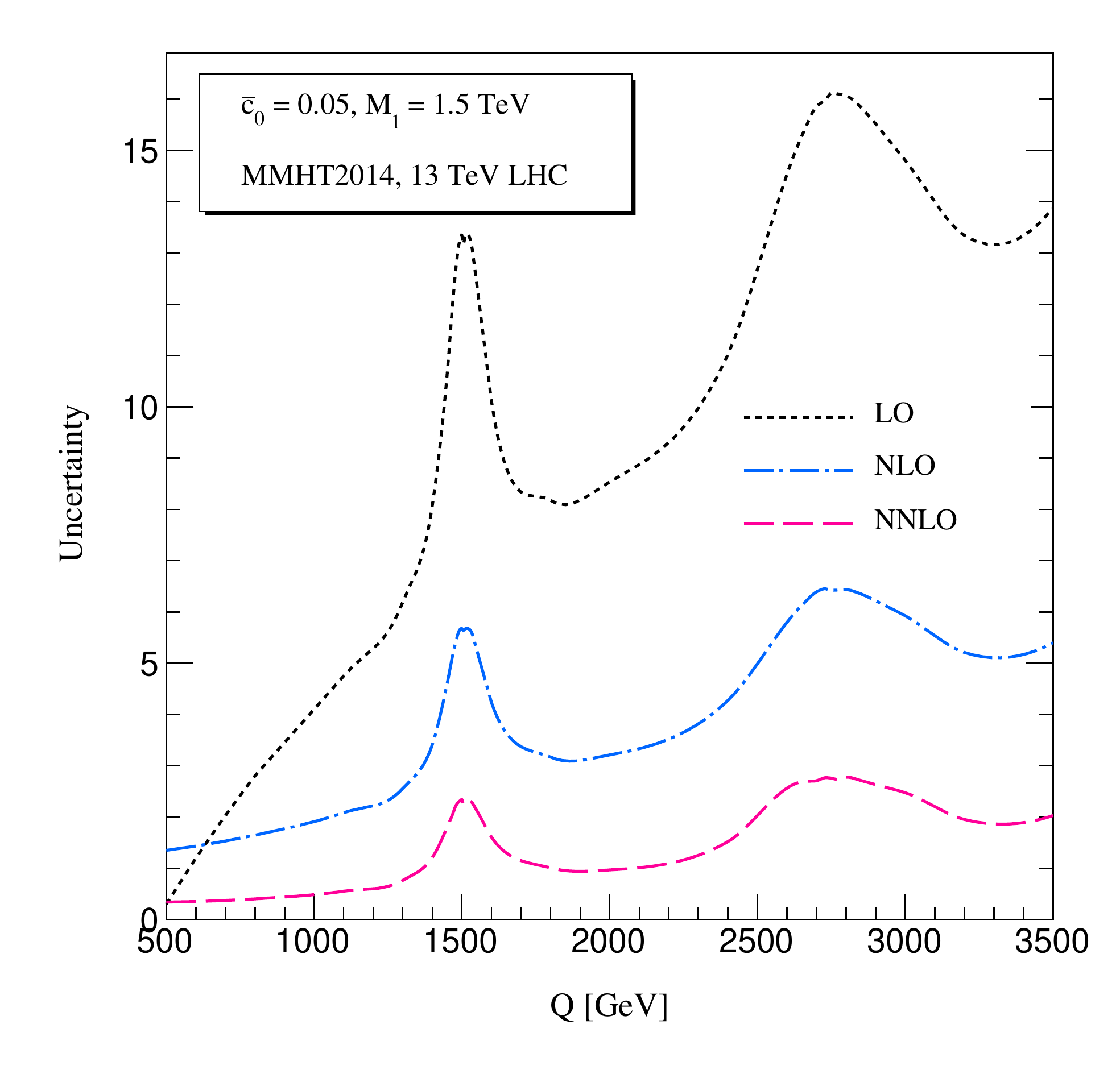}}

\caption{$7$-point scale variation in the signal is shown up to NNLO for the di-lepton invariant mass distribution.} 
\label{nnlo_scale}
\end{figure}

We have considered different sources of theoretical uncertainties in our analysis. First, we considered the uncertainties due to the presence of two unphysical scales $\mu_r$ and $\mu_f$ in the theory and then those coming from the non-perturbative parton distribution function in the calculation. For the scale uncertainties we vary $\mu_r$ and $\mu_f$ simultaneously from $Q/2$ to $2Q$ by putting the constraint that the ratio of unphysical scales is less than $2$, as
\begin{align}
        \Big|\text{ln}\frac{\mu_r}{Q}\Big| \leq \text{ln }2, \quad \Big|\text{ln}\frac{\mu_f}{Q}\Big| \leq \text{ln }2,
        \quad \Big|\text{ln}\frac{\mu_r}{\mu_f}\Big| \leq \text{ln }2.
        \label{eqscale}
\end{align}
The last condition in \eq{eqscale} ensures that no unusual choice of the scale combination is considered within the range. This results in $7$ different combinations of the scale \viz $\left(\mu_r/Q, \mu_f/Q \right) = (1/2, 1/2), (1/2, 1), (1, 1/2)$, $(1, 1), (2, 1), (1, 2), (2, 2)$. With this choice, we estimate the $7$-point scale uncertainties in the di-lepton invariant mass distribution up to NNLO and the results are depicted in \fig{nnlo_scale}. The scale uncertainties are found to get reduced significantly from LO to NNLO over the full invariant mass region. For $Q = 1500$ GeV \ie right at the first resonance, the scale uncertainties at LO are $\pm{13.3\%}$, at NLO they are $\pm{5.7\%}$, at NNLO it further reduces to $\pm{2.3\%}$. Away from resonance, the uncertainty also decreases order by order. For example at $Q = 3000$ GeV, the scale uncertainties get reduced from $\pm{14.8\%}$ at LO to about $\pm{2.5\%}$ at NNLO. In the off-resonance region the uncertainty in general increases with increasing $Q$ which can be tamed with inclusion of further higher order terms in the perturbation theory.

%
\input{table3}
We also estimate the uncertainties coming from the non-perturbative PDFs. For this we calculate the uncertainty due to the intrinsic errors in the PDFs that result from various experimental errors from the global fits. In this cases we use the PDF sets {\tt ABMP16} \cite{Alekhin:2013nda}, {\tt CT14} \cite{Dulat:2015mca}, {\tt MMHT2014} \cite{Harland-Lang:2014zoa}, {\tt NNPDF31} \cite{Ball:2017nwa},  and {\tt PDF4LHC15} \cite{Butterworth:2015oua} provided from the {\tt lhapdf}. The central predictions for these different PDF groups also differ due to different underlying assumptions in global fits for different groups.  We calculate the intrinsic PDF uncertainties using $51$ sets for {\tt MMHT2014}, $57$ sets for {\tt CT14}, $101$ sets for {\tt NNPDF31}, $30$ sets for {\tt ABMP16} and $31$ sets for {\tt PDF4LHC15}. To this end we use all PDF sets extracted at NNLO level. In \tab{table3} we present these uncertainties for the di-lepton invariant mass distribution to NNLO.  We find that around the resonance $M_1 =1500$ the PDF uncertainty is well within $5\%$ except for the {\tt CT14} which shows relatively increased uncertainty. In the high invariant mass region the uncertainty however increases due to unavailability of sufficient data in those region.

\begin{figure}
  \centering
                   {\includegraphics[trim=0 0 0 0,clip,width=0.48\textwidth,height=0.40\textwidth]
                   {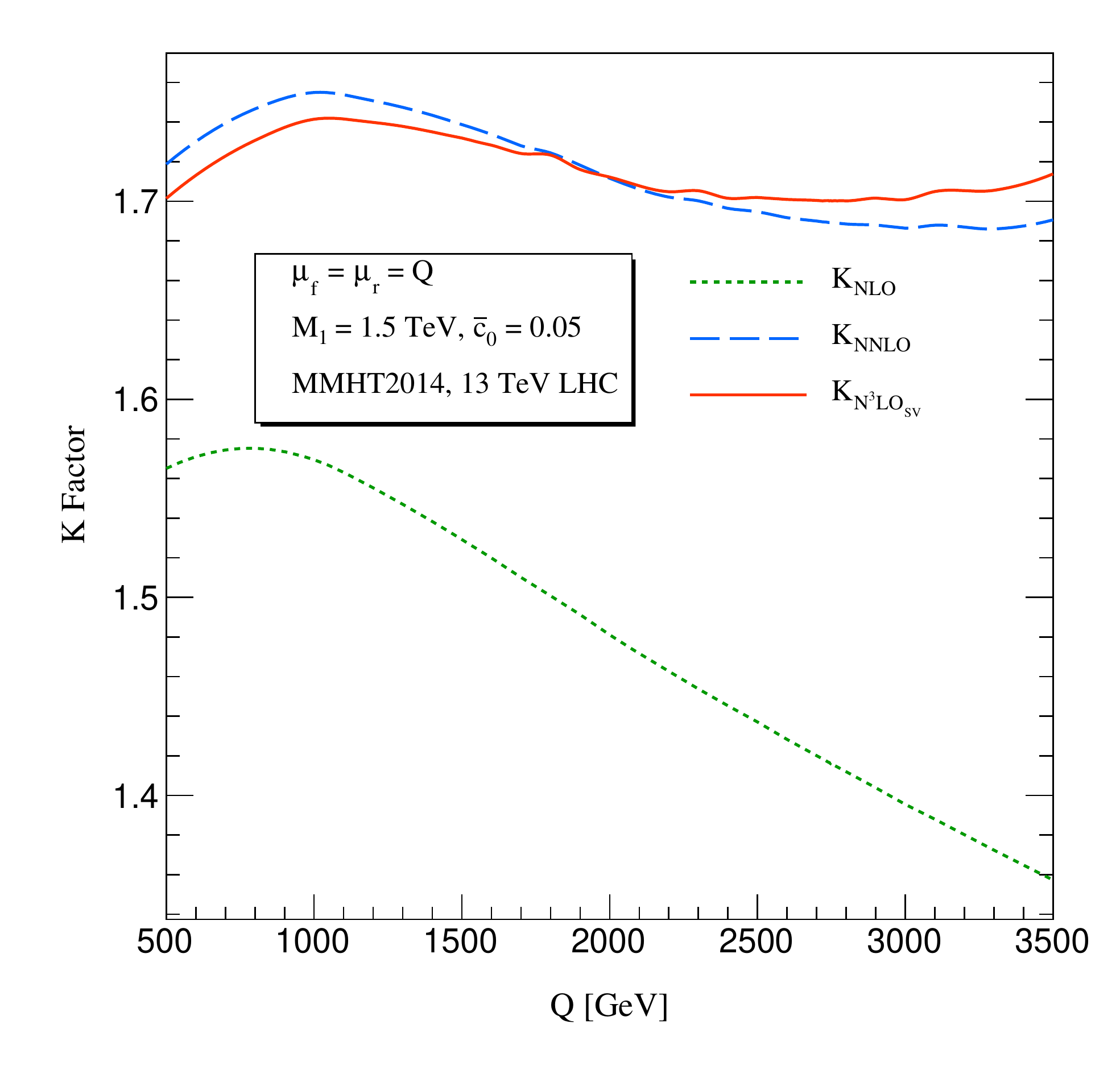}}
  \hskip0.05cm
                   {\includegraphics[trim=0 0 0 0,clip,width=0.48\textwidth,height=0.40\textwidth]
                   {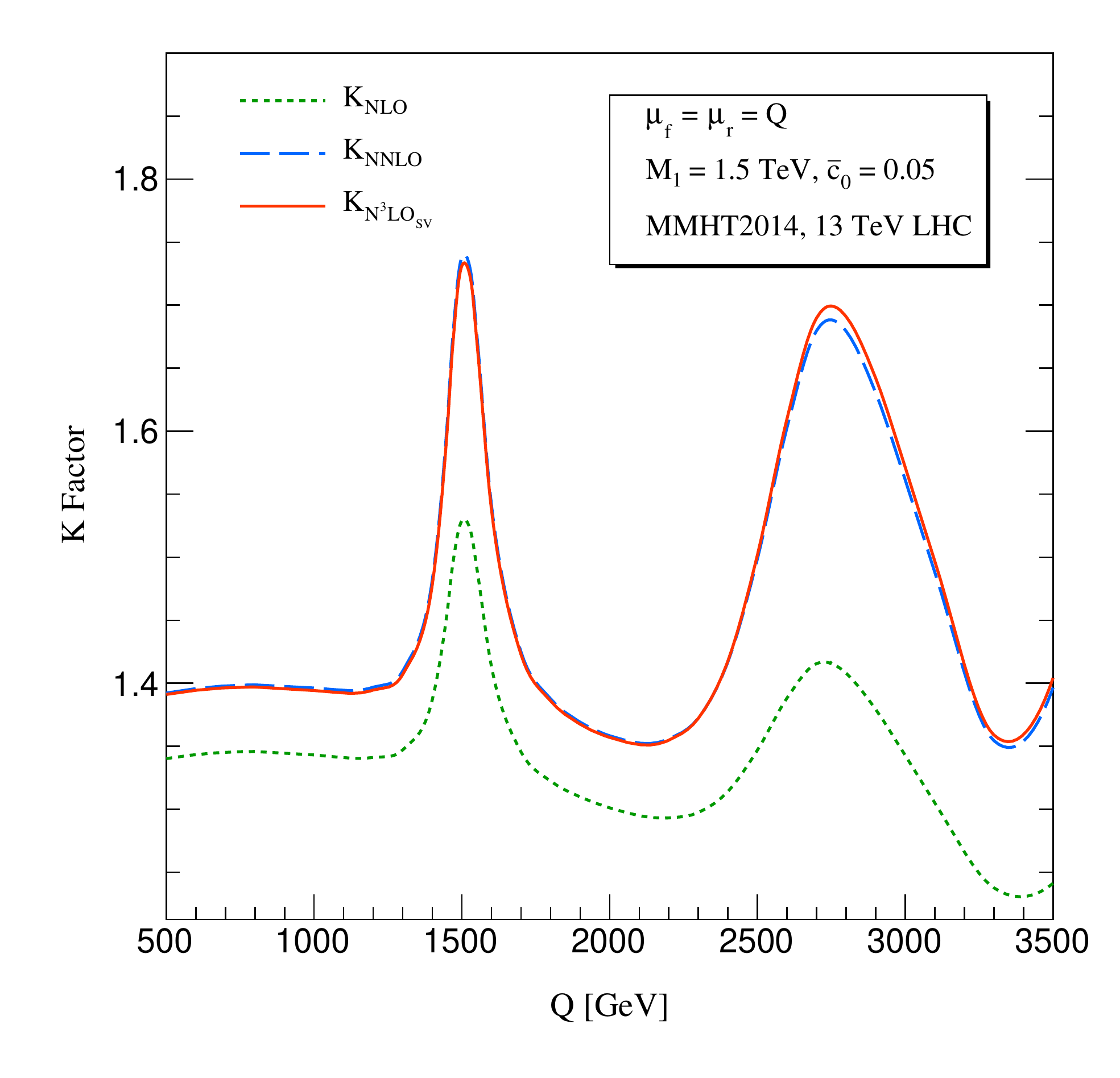}}

	\caption{K-factors are presented up to N$^3$LO$_{sv}$ for the RS model (left) and for the signal
        (right).}
\label{n3lo_kfac}
\end{figure}
From the above observation we notice that the NNLO corrections for RS production are large enough to truncate the perturbation theory at this order and necessitates the computation of higher order corrections for the convergence of the perturbation series. As a first step beyond NNLO we studied the three-loop SV correction for di-lepton production channel using the universal property of the SV coefficients for generic spin-2 couplings. In \fig{n3lo_kfac} we present these three loop SV corrections in terms of the corresponding $K$-factors up to N$^3$LO$_{\rm sv}$ as a function of the di-lepton invariant mass for pure RS case (left panel) as well as for the signal (right panel).  We use {\tt MMHT2014nnlo} set for this analysis.
These three-loop SV corrections are found to contribute an additional -0.7\% of LO to the NNLO result at first resonance $M_1 =1500$ GeV for pure RS case,  demonstrating a very good convergence of the perturbation theory at this order. In the high invariant mass region away from the RS resonance however we see correction due to third order SV terms is about $1\%$ of LO cross-section.
 We also note that the three-loop SV corrections are negative in the low $Q$-region while in the high $Q$-region they are positive because of threshold enhancement.  The 7-point scale uncertainty  is seen to increase at the lower invariant mass region whereas as we reach higher invariant mass region this becomes better. To further constraint the scale uncertainty the N$^3$LO PDFs are essential at this order. Also the missing sub-leading pieces are important in particular in the low-$Q$ region (see $eg.$ \cite{deFlorian:2014vta,Das:2020adl}). The $\mu_r$ uncertainty however is seen to improve in the whole invariant mass region. Keeping $\mu_f=Q=M_1$ we observe an uncertainty of $\pm 0.9\%$ around the resonance $M_1=1500$ for the variation in the range $(1/2,2)M_1$.

\subsection{Resum result}\label{sec:resummation}
\begin{figure}[h!]
  \centering

                   {\includegraphics[trim=0 0 0 0,clip,width=0.48\textwidth,height=0.40\textwidth]
                   {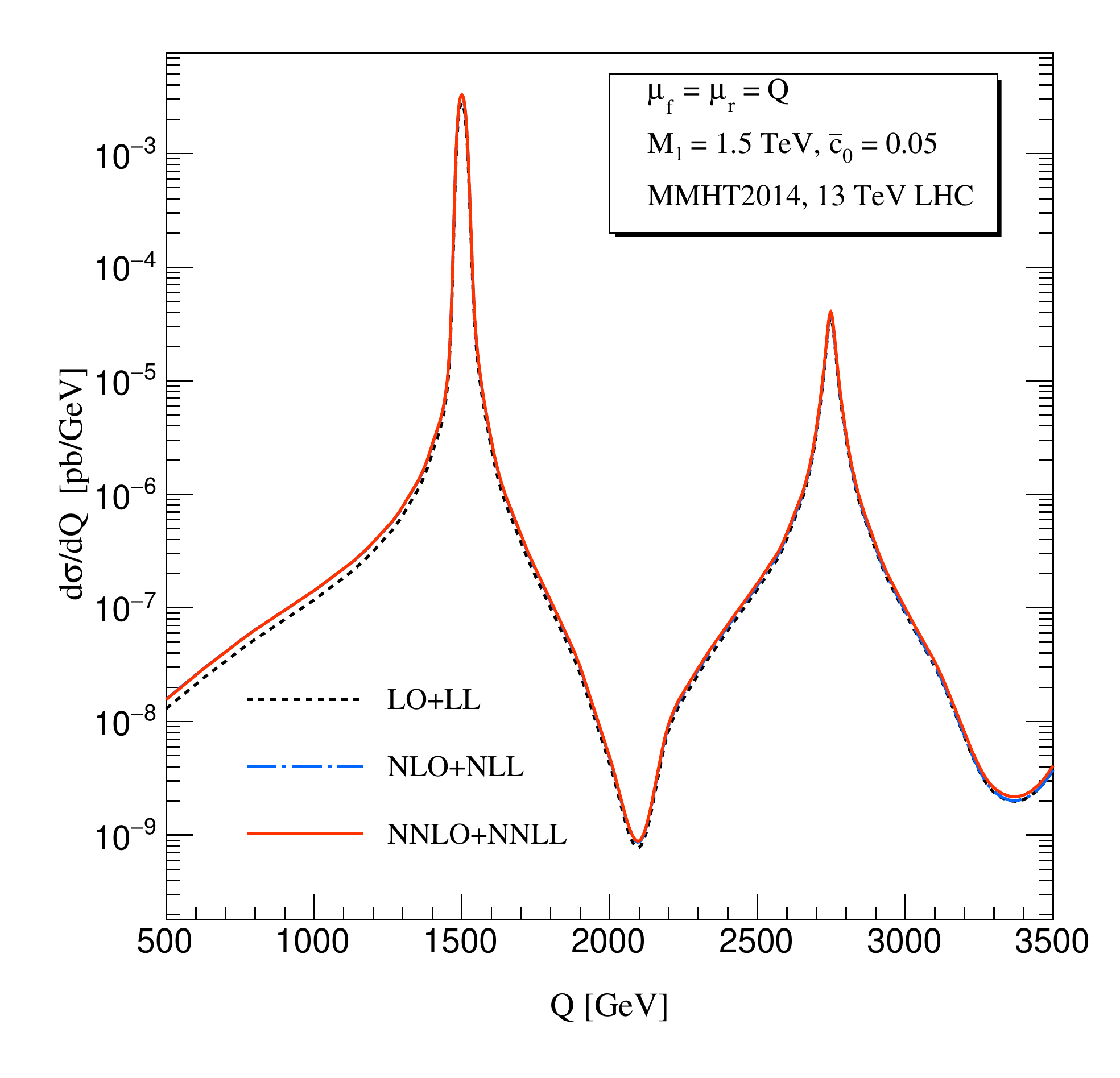}}
  \hskip0.05cm
                   {\includegraphics[trim=0 0 0 0,clip,width=0.48\textwidth,height=0.40\textwidth]
                   {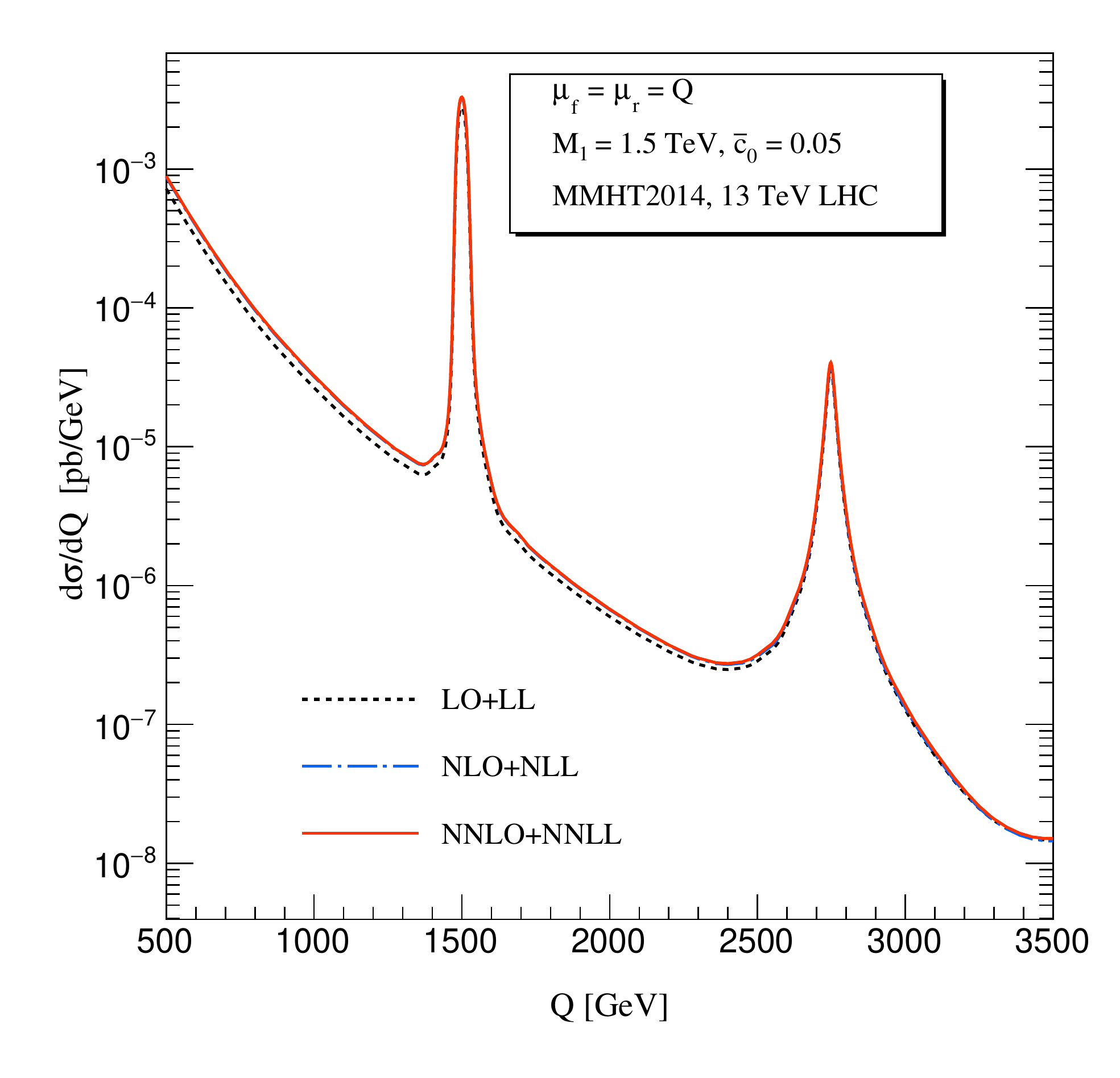}}

        \caption{Di-lepton invariant mass distribution up to NNLO+NNLL for RS model
        (left) and for the signal (right).}
\label{res_inv_distribution}
\end{figure}

We now move to study the effect of threshold logarithms by resumming them to NNLL accuracy and match to the computed NNLO cross-section in the \sect{sec:fo}. For this, the same choice of SM and RS model parameters has been used as in the fixed order computation. For the inverse Mellin transformation \eq{eq:matched}, we use $c = 1.9$.  In \fig{res_inv_distribution} we present the di-lepton invariant mass  distribution for GR and for the signal at different logarithmic accuracy. 

\begin{figure}[h!]
  \centering

                   {\includegraphics[trim=0 0 0 0,clip,width=0.48\textwidth,height=0.40\textwidth]
                   {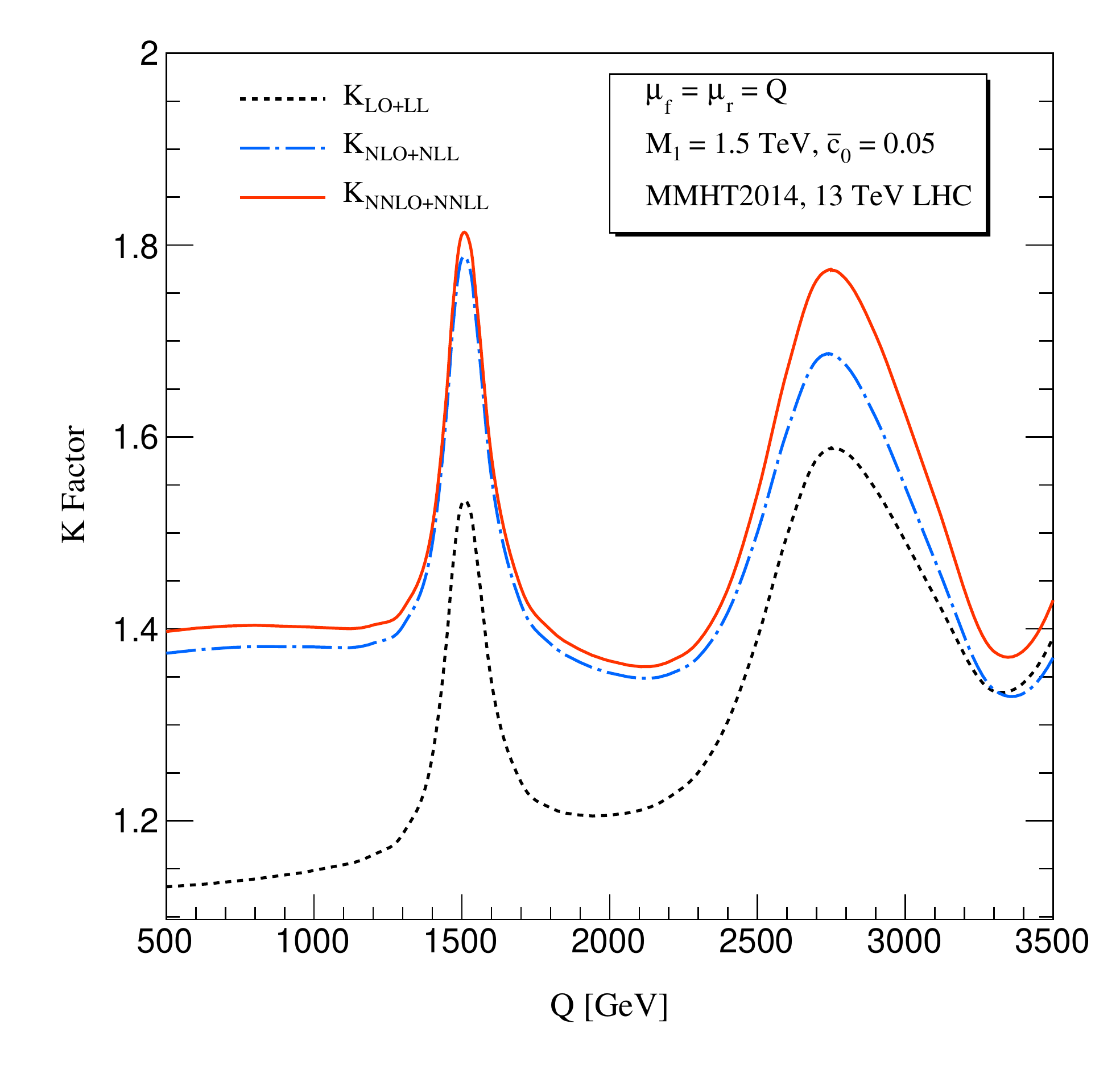}}
  \hskip0.05cm
                   {\includegraphics[trim=0 0 0 0,clip,width=0.48\textwidth,height=0.40\textwidth]
                   {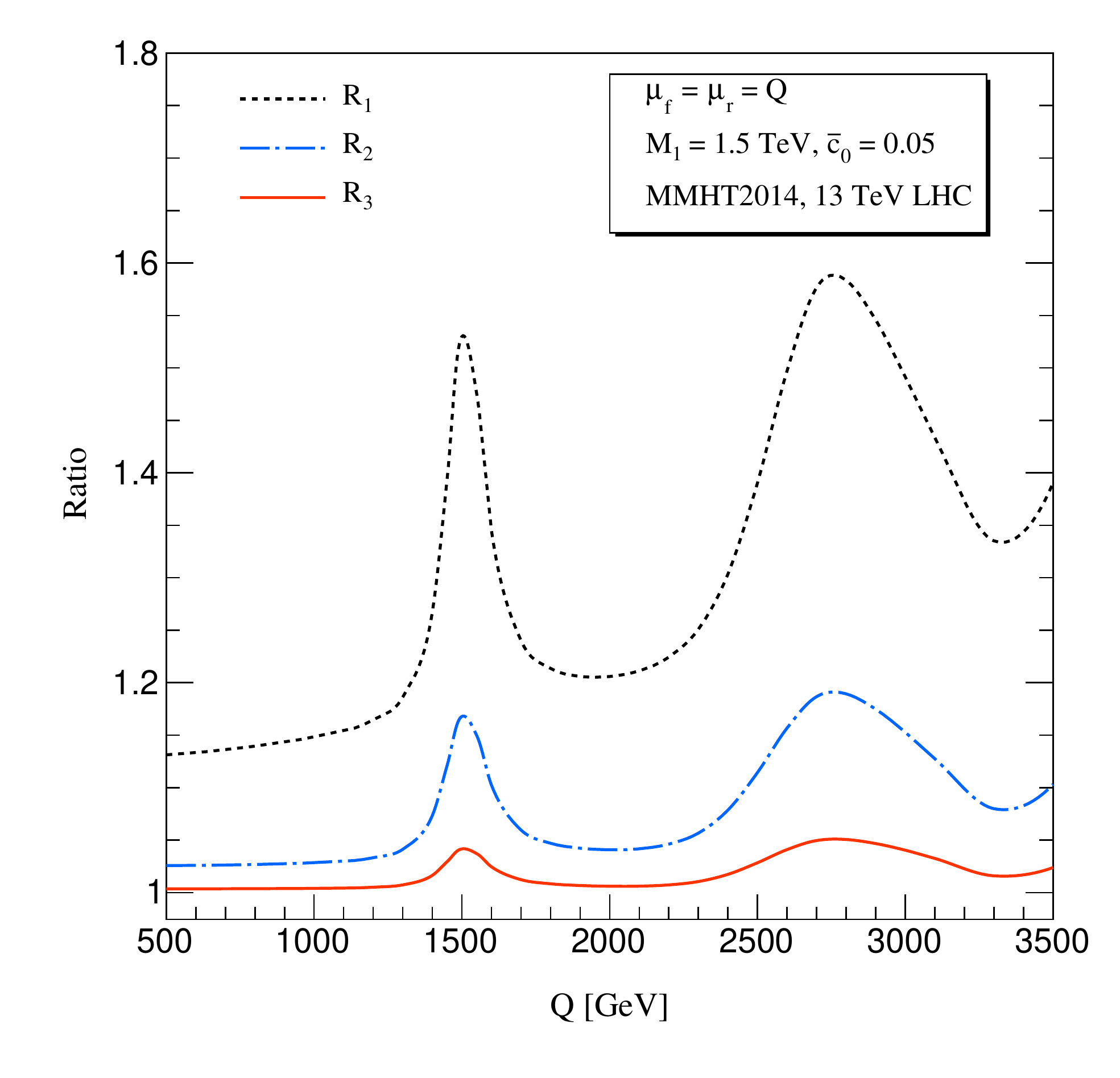}}

        \caption{Resummed K-factors for the  di-lepton invariant mass distribution as defined in \eq{resum_k}
and the corresponding ratios as defined in \eq{resum_ratio} (right).}
\label{res_k_distribution}
\end{figure}

\begin{align}\label{resum_k}
        \text{K}_{\rm LL} = \frac{ d\sigma^{\rm LO+LL}/dQ} {d\sigma^{\rm LO}/dQ }  \,, ~
         \text{K}_{\rm NLL} = \frac{ d\sigma^{\rm NLO+NLL}/dQ} {d\sigma^{\rm LO}/dQ } \,,~
        \text{K}_{\rm NNLL} = \frac{ d\sigma^{\text{N$^2$LO+N$^2$LL}}/dQ} {d\sigma^{\rm LO}/dQ } \,.
\end{align}
%
\input{table4}
To quantify these resummation effects, we define the resum K-factors in \eq{resum_k} and present the same in \tab{table4} for different $Q$ values. The enhancement due to threshold logarithms for the signal is significant for all $Q$ values, however it is more significant at the resonance region. This is because of the underlying born processes for the graviton production in the RS model. At the born level, the RS graviton can be produced via quark-antiquark annihilation process (DY-like) as well as gluon fusion channel (Higgs-like). It is well known that the QCD corrections, particularly, the threshold enhancement in these two channels are different and are more pronounced for gluon fusion channel. Here, the signal receives contribution from RS (DY-like as well as Higgs-like) and the SM background (DY-like). However, at the resonance region GR dominates over the SM background by several orders of magnitude and hence the threshold enhancement due to the gluon fusion channel becomes prominent. Far off the resonance region, the signal is essentially dominated by the SM background and assumes DY-like threshold enhancement. For completeness, we present these resummed K-factors for GR case in \fig{res_k_distribution}. 
\begin{align}\label{resum_ratio}
	 \text{R}_{2} = \frac{ d\sigma^{\rm NLO+NLL}/dQ} {d\sigma^{\rm NLO}/dQ } \,,~
        \text{R}_{3} = \frac{ d\sigma^{\text{N$^2$LO+N$^2$LL}}/dQ} {d\sigma^{\rm NNLO}/dQ } \,.
\end{align}
In order to further study the enhancement due to threshold resummation for the signal, we consider the ratios of the resummed results to the fixed order results defined in \eq{resum_ratio}. We observe that at resonance ($Q=M_1=1500$ GeV), NNLO+NNLL contributes additional $4\%$ enhancement over NNLO. These ratios are presented in \fig{res_k_distribution}. Moreover, in \tab{table5}, we present these resum $K$-factors right at the resonance for different values of resonance mass $M_1$.

%
\input{table5.tex}

\begin{figure}[h!]
  \centering

                   {\includegraphics[trim=0 0 0 0,clip,width=0.60\textwidth,height=0.50\textwidth]
                   {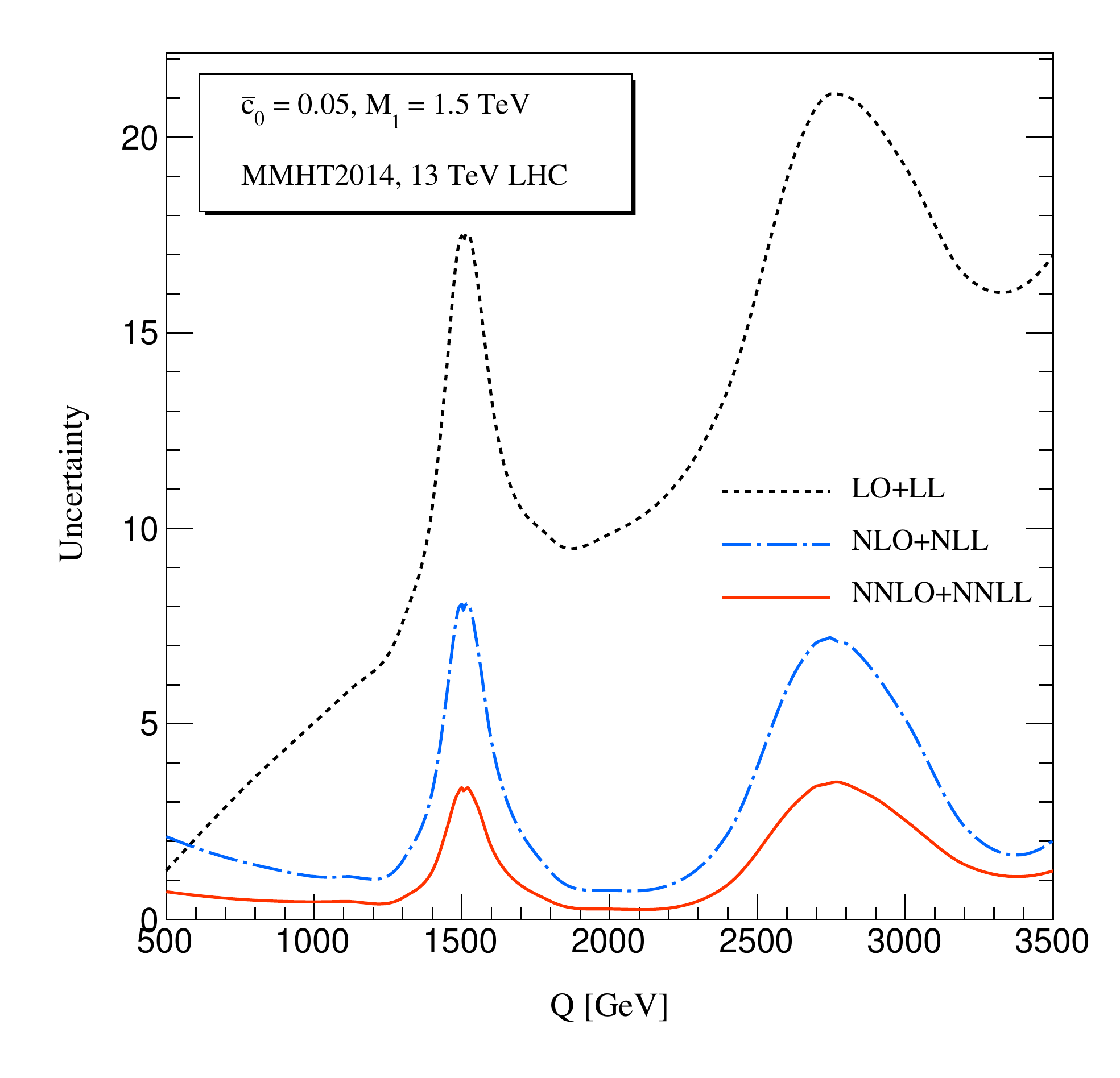}}

\caption{$7$-point scale uncertainties in the signal are shown up to NNLO+NNLL for di-lepton invariant mass distribution.}
\label{res_scale}
\end{figure}

Next, we estimate the theoretical uncertainties in resummed predictions  due to the unphysical scales
$\mu_r$ and $\mu_f$ as well as due to the non-perturbative PDFs. The conventional $7$-point scale uncertainties for
the signal are presented in \fig{res_scale} for different logarithmic accuracy. At the resonance region, these scale uncertainties are estimated to be about $\pm 17.5\%$, $\pm 8.1\%$ and $\pm 3.4\%$ at LO+LL, NLO+NLL and NNLO+NNLL respectively. 
Moreover, these uncertainties are bit larger than the corresponding ones for the fixed order results presented
in \fig{nnlo_scale}.

\begin{figure}[h!]
  \centering

                   {\includegraphics[trim=0 0 0 0,clip,width=0.48\textwidth,height=0.40\textwidth]
                   {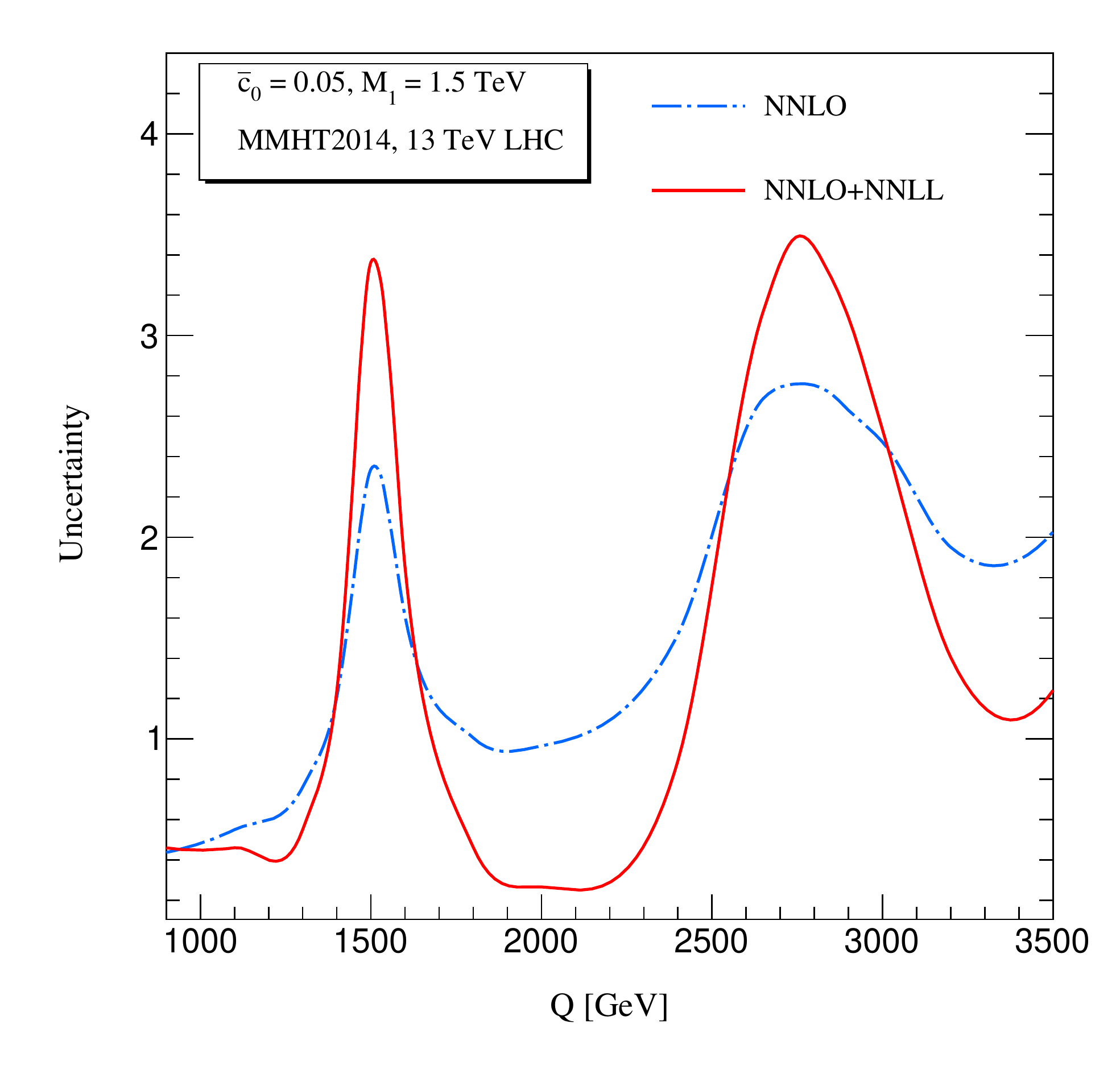}}
  \hskip0.05cm
                   {\includegraphics[trim=0 0 0 0,clip,width=0.48\textwidth,height=0.40\textwidth]
                   {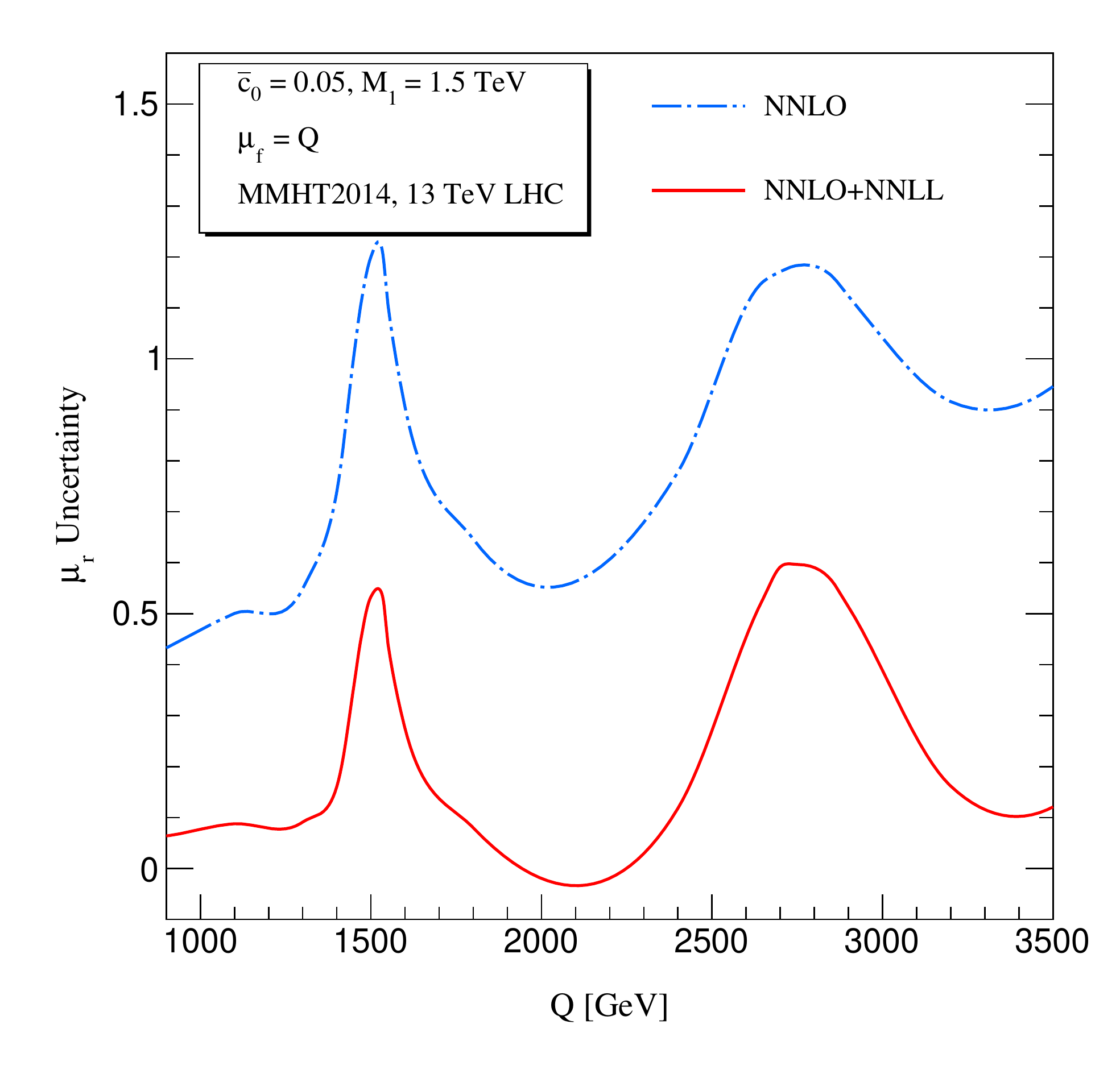}}		   

	\caption{Comparison of $7$-point scale uncertainties at the signal for NNLO and NNLO+NNLL (left). 
The uncertainty only due to $\mu_r$ scale variation around the central scale  $Q$ at NNLO and NNLO+NNLL (right)
for fixed $\mu_f=Q$.}
\label{scale_resum_vs_fo}
\end{figure}

To estimate these uncertainties at the NNLO level and beyond, we contrast these scale uncertainties
in the resummed results against those in the fixed order results in \fig{scale_resum_vs_fo} (left panel).
For the resummed case, the scale uncertainty at NNLO+NNLL for $Q=M_1$ is about $3.4\%$ and is larger 
than the one $2.3\%$ at NNLO level. 
This increase in scale uncertainty can be understood from the fact that in the resummation formalism only threshold
logarithms that are significant in the limit $z \to 1$ have been resummed to all orders in QCD but not the logarithms of
unphysical scales. Moreover, it is observed that for Higgs-like processes  the resummation does not improve the
scale uncertainties over the fixed order ones \cite{Bonvini:2014joa} for any choice of central scales. 
In the present context, the graviton production at the resonance
receives significant contribution from this Higgs-like gluon fusion process and hence the associated large scale uncertainty.
However, the scale uncertainties only due to the renormalization scale $\mu_r$ are found to get reduced from fixed order
NNLO level $\pm 1.2\%$ at $Q=1500$ to the resummed NNLO+NNLL level $\pm 0.5\%$  ( see right panel of \fig{scale_resum_vs_fo}). 

Further, we estimate in our predictions the uncertainty due to the non-perturbative PDF inputs. These uncertainties
are obtained for each PDF group by systematically calculating the cross section for each of the available sets. These 
PDF uncertainties are presented for different resonance mass $M_1$ values in \tab{table6}. This uncertainty 
for the kinematic range considered and the PDF groups studied, is smallest at $M_1=1500$ GeV for {\tt NNPDF31} and largest
for {\tt CT14} at $M_1=3500$ GeV. 
%
\input{table6}
%

\section{Conclusions}\label{sec:conclusion}
In the absence of any signature of new physics at the LHC, it is high time to explore possible scenarios where we could make potential discovery of new physics beyond the SM. 
In particular the RS model provides to be a very good candidate in the search of massive spin-2 resonances. In the literature, it is 
found that the NLO QCD corrections to this process are quite substantial in the di-lepton channel, implying the need for higher order corrections for
giving precise theory predictions that augment the search for RS gravitons at the collider experiments.
In this work, we have studied the NNLO QCD corrections for the di-lepton production process through graviton propagator and 
have presented for results for the di-lepton invariant mass distribution up to $Q$ values as high as $3.5$ TeV. 
The underlying born contributions for this process receive both DY-like as well as Higgs-like contributions and hence the 
corresponding QCD corrections for the signal at the resonance region are very significant,  while the QCD corrections off the resonance are mostly SM DY-like. This results in K-factors that are strongly dependent on the invariant mass
of the di-lepton. We have presented these mass dependent K-factors at NNLO and beyond for 13 TeV LHC.
We find that while NLO correction is about $53\%$ of LO, the NNLO correction increases the cross section by
additional $21\%$.
The scale uncertainty in the NNLO result at the resonance region also got significantly reduced to as small as $2\%$ 
for $Q=M_1=1500$ GeV.

Further, we have extended our work to include the important SV corrections at the N$^3$LO level.
We find that the SV contribution at this order for $Q=1500$ GeV is about $0.7\%$ of LO in magnitude but negative in sign, thus
demonstrating a very good convergence of the perturbation series.
In addition we also studied the threshold resummation by resumming all the large-threshold logarithms to NNLL accuracy. 
We have presented these results by matching the NNLL resummed results to the fixed order NNLO ones. 
We find that these resummed results contribute an additional $7\%$ of LO to the NNLO ones.
To conclude, we note that our results are most precise theoretical predictions available to date and
that these mass dependent K-factors will be useful in the search for RS graviton resonances
in the experimental data analysis using di-lepton events at the LHC.
\section*{Acknowledgements}
The authors would like to thank V. Ravindran for useful discussion. The research of G.D. is supported by the \textit{Deutsche Forschungsgemeinschaft} (DFG) within the Collaborative Research Center TRR 257 (\textit{Particle Physics Phenomenology after the Higgs Discovery}).

\bibliographystyle{JHEP}
\bibliography{references}
\end{document}

%% file: table1.tex
\begin{table}
	\begin{center}	
	 \begin{tabular}{|c|c|c|}
\hline
	$Q$ (GeV) & ${\rm K}_{\rm NLO}$  & ${\rm K}_{\rm NNLO}$ \\

\hline
	$500$&  $1.340$  & $1.392$  \\
\hline
        $1000$& $1.343$  & $1.396$  \\
\hline
        $1500$& $1.529$  & $1.738$  \\
\hline
        $2000$& $1.301$  & $1.358$  \\
\hline
        $2500$& $1.347$  & $1.499$  \\
\hline
\end{tabular}
	\caption{\small{The fixed order K-factors for the signal up to NNLO in QCD for $13$ TeV LHC.}}
\label{table1}
 \end{center}
\end{table}

%% file: table2.tex
\begin{table}
	\begin{center}	
	 \begin{tabular}{|c|c|c|}
\hline
	M$_1$ (GeV) & ${\rm K}_{\rm NLO}$  & ${\rm K}_{\rm NNLO}$ \\

\hline
	$1500$&  $1.529$  & $1.738$ \\
\hline
        $2000$& $1.481$  & $1.713$  \\
\hline
        $2500$& $1.436$  & $1.694$  \\
\hline
        $3000$& $1.395$  & $1.686$  \\
\hline
        $3500$& $1.357$  & $1.694$  \\
\hline
        $4000$& $1.320$  & $1.718$  \\
\hline
        $4500$& $1.283$  & $1.760$  \\
\hline
\end{tabular}
	\caption{\small{The fixed order K-factors for the signal to NNLO in QCD  right at the resonance region 
for different $M_1$ values are presented for $13$ TeV LHC.}}
\label{table2}
 \end{center}
\end{table}

%% file: table3.tex
\begin{table}[h!]
	\begin{center}
{\scriptsize
		\resizebox{15.1cm}{1.2cm}{
	 		\begin{tabular}{|c|c|c|c|c|c|c|}
\hline
		 M$_1$ (GeV) & {\tt ABMP16 } & {\tt CT14} & {\tt MMHT2014} &  {\tt NNPDF31}  & {\tt PDF4LHC15} \\

\hline
$1500$& $2.66\times10^{-3} (\pm 4.7\%)$ & $3.13\times 10^{-3}(\pm 9.6\%)$ &$3.17\times 10^{-3}(\pm 4.0\%)$ 
& $2.91\times 10^{-3}(\pm 2.1\%)$ & $3.11\times 10^{-3}(\pm 5.0\%)$ \\
\hline
$2000$& $3.35\times 10^{-4}(\pm 5.7\%)$ & $4.06\times 10^{-4}(\pm 12.0\%)$ & $4.20\times 10^{-4}(\pm 5.0\%)$ &
$3.72\times 10^{-4}(\pm 2.6\%)$ &  $4.07\times 10^{-4}(\pm 6.5\%)$  \\
\hline
$2500$& $5.84\times 10^{-5}(\pm 6.0\%)$ & $7.25\times 10^{-5}(\pm 14.4\%)$ & $7.70\times 10^{-5}(\pm 5.9\%)$ &
$6.53\times 10^{-5}(\pm 3.4\%)$ &  $7.35\times 10^{-5}(\pm 7.9\%)$  \\
\hline
$3000$& $1.23\times 10^{-5}(\pm 6.3\%)$ & $1.56\times 10^{-5}(\pm 17.0\%)$ & $1.71\times 10^{-5}(\pm 6.9\%)$ &
$1.38\times 10^{-5}(\pm 5.8\%)$ & $1.61\times 10^{-5}(\pm 9.5\%)$  \\
\hline
$3500$& $2.91\times 10^{-6}(\pm 6.5\%)$ & $3.78\times 10^{-6}(\pm 20.0\%)$ & $4.26\times 10^{-6}(\pm 8.0\%)$ &
$3.20\times 10^{-6}(\pm 11.9\%)$ & $3.96\times 10^{-6}(\pm 11.2\%)$  \\
\hline
\end{tabular}
 }
		\caption{\small{Intrinsic PDF uncertainties in the signal at NNLO QCD for different PDF choices 
are given right at the resonance for different $M_1$ values.
All the results are presented for $13$ TeV LHC. The cross sections are given for the central set ($n=0$) for each
PDF group along with the corresponding intrinsic uncertainties in terms of the percentage.}} 
\label{table3}
 }
 \end{center}
\end{table}

%% file: table4.tex
\begin{table}
	\begin{center}	
	 \begin{tabular}{|c|c|c|c|c|c|}
\hline
		 $Q$ (GeV) & ${\rm K}_{\rm LL}$  & ${\rm K}_{\rm NLL}$ & ${\rm K}_{\rm NNLL}$ & R$_2$ & R$_3$  \\

\hline
	$500$&  $1.131$  & $1.375$ & $1.397$ & $1.026$ & $1.004$  \\
\hline
        $1000$& $1.148$  & $1.381$ & $1.401$ & $1.029$ & $1.004$ \\
\hline
        $1500$& $1.530$  & $1.786$ & $1.811$ & $1.168$ & $1.042$\\
\hline
        $2000$& $1.206$  & $1.354$ & $1.367$ & $1.041$ & $1.006$ \\
\hline
        $2500$& $1.390$  & $1.501$ & $1.541$ & $1.114$ & $1.028$ \\
\hline
\end{tabular}
	\caption{\small{Resum K-factors and the ratios as defined in \eq{resum_k} and \eq{resum_ratio} as a function of 
the di-lepton invariant mass for the default choice of RS model parameters.}}
\label{table4}
 \end{center}
\end{table}

%% file: table5.tex
\begin{table}
	\begin{center}	
	 \begin{tabular}{|c|c|c|c|}
\hline
	M$_1$ (GeV) & ${\rm K}_{\rm LL}$  & ${\rm K}_{\rm NLL}$ & ${\rm K}_{\rm NNLL}$ \\

\hline
	$1500$ & $1.530$  & $1.786$ & $1.811$ \\
\hline
        $2000$ & $1.550$  & $1.743$ & $1.790$ \\
\hline
        $2500$ & $1.574$  & $1.703$ & $1.776$ \\
\hline
        $3000$ & $1.603$  & $1.670$ & $1.777$ \\
\hline
        $3500$ & $1.639$  & $1.644$ & $1.793$ \\
\hline
        $4000$ & $1.681$  & $1.623$ & $1.832$ \\
\hline
        $4500$ & $1.730$  & $1.607$ & $1.891$ \\
\hline
\end{tabular}
	\caption{\small{Resum K-factors for signal right at the resonance for different $M_1$ values are presented
up to NNLO+NNLL in QCD for $13$ TeV LHC.}}
\label{table5}
 \end{center}
\end{table}

%% file: table6.tex
\begin{table}[h!]
	\begin{center}
{\scriptsize		
\resizebox{15.1cm}{1.2cm}{
		\begin{tabular}{|c|c|c|c|c|c|c|}
\hline
		 M$_1$ (GeV) & {\tt ABMP16}  & {\tt CT14} & {\tt MMHT2014} & {\tt NNPDF31} & {\tt PDF4LHC15} \\

\hline
$1500$& $2.77\times 10^{-3}(\pm 4.7\%)$ & $3.26\times 10^{-3}(\pm 9.4\%)$ & $3.30\times 10^{-3}(\pm 3.8\%)$ &
$3.04\times 10^{-3}(\pm 2.1\%)$ & $3.25\times 10^{-3}(\pm 4.9\%)$ \\
\hline
$2000$& $3.50\times 10^{-4}(\pm 5.7\%)$ & $4.25\times 10^{-4}(\pm 11.9\%)$ & $4.39\times 10^{-4}(\pm 4.7\%)$ &
$3.90\times 10^{-4}(\pm 2.6\%)$ & $4.26\times 10^{-4}(\pm 6.4\%)$  \\
\hline
$2500$& $6.12\times 10^{-5}(\pm 6.1\%)$ & $7.62\times 10^{-5}(\pm 14.3\%)$ & $8.06\times 10^{-5}(\pm 5.6\%)$ &
$6.87\times 10^{-5}(\pm 3.3\%)$ & $7.72\times 10^{-5}(\pm 7.8\%)$  \\
\hline
$3000$& $1.30\times 10^{-5}(\pm 6.4\%)$ & $1.65\times 10^{-5}(\pm 16.9\%)$ & $1.79\times 10^{-5}(\pm 6.5\%)$ &
$1.45\times 10^{-5}(\pm 5.6\%)$ & $1.69\times 10^{-5}(\pm 9.4\%)$  \\
\hline
$3500$& $3.08\times 10^{-6}(\pm 6.7\%)$ & $4.02\times 10^{-6}(\pm 19.8\%)$ & $4.49\times 10^{-6}(\pm 7.5\%)$ &
$3.41\times 10^{-6}(\pm 11.2\%)$ & $4.20\times 10^{-6}(\pm 11.1\%)$  \\
\hline
\end{tabular}
 }
		\caption{\small{Intrinsic PDF uncertainties in the signal at NNLO+NNLL QCD for different PDF choices 
are given right at the resonance for different $M_1$ values.
All the results are presented for $13$ TeV LHC. The cross sections are given in terms of pb for the central set ($n=0$) for each
PDF group along with the corresponding intrinsic uncertainties in terms of the percentage.}} 
\label{table6}
 }
 \end{center}
\end{table}

%% file: rs_resum.bbl
\providecommand{\href}[2]{#2}\begingroup\raggedright\begin{thebibliography}{100}

\bibitem{Aad:2012tfa}
{\scshape ATLAS} collaboration, G.~Aad et~al., \emph{{Observation of a new
  particle in the search for the Standard Model Higgs boson with the ATLAS
  detector at the LHC}},
  \href{https://doi.org/10.1016/j.physletb.2012.08.020}{\emph{Phys. Lett.}
  {\bfseries B716} (2012) 1} [\href{https://arxiv.org/abs/1207.7214}{{\ttfamily
  1207.7214}}].

\bibitem{Chatrchyan:2012xdj}
{\scshape CMS} collaboration, S.~Chatrchyan et~al., \emph{{Observation of a new
  boson at a mass of 125 GeV with the CMS experiment at the LHC}},
  \href{https://doi.org/10.1016/j.physletb.2012.08.021}{\emph{Phys. Lett.}
  {\bfseries B716} (2012) 30}
  [\href{https://arxiv.org/abs/1207.7235}{{\ttfamily 1207.7235}}].

\bibitem{Randall:1999ee}
L.~Randall and R.~Sundrum, \emph{{A Large mass hierarchy from a small extra
  dimension}}, \href{https://doi.org/10.1103/PhysRevLett.83.3370}{\emph{Phys.
  Rev. Lett.} {\bfseries 83} (1999) 3370}
  [\href{https://arxiv.org/abs/hep-ph/9905221}{{\ttfamily hep-ph/9905221}}].

\bibitem{Harlander:2002wh}
R.~V. Harlander and W.~B. Kilgore, \emph{{Next-to-next-to-leading order Higgs
  production at hadron colliders}},
  \href{https://doi.org/10.1103/PhysRevLett.88.201801}{\emph{Phys. Rev. Lett.}
  {\bfseries 88} (2002) 201801}
  [\href{https://arxiv.org/abs/hep-ph/0201206}{{\ttfamily hep-ph/0201206}}].

\bibitem{Harlander:2002vv}
R.~V. Harlander and W.~B. Kilgore, \emph{{Production of a pseudoscalar Higgs
  boson at hadron colliders at next-to-next-to leading order}},
  \href{https://doi.org/10.1088/1126-6708/2002/10/017}{\emph{JHEP} {\bfseries
  10} (2002) 017} [\href{https://arxiv.org/abs/hep-ph/0208096}{{\ttfamily
  hep-ph/0208096}}].

\bibitem{Anastasiou:2002yz}
C.~Anastasiou and K.~Melnikov, \emph{{Higgs boson production at hadron
  colliders in NNLO QCD}},
  \href{https://doi.org/10.1016/S0550-3213(02)00837-4}{\emph{Nucl. Phys.}
  {\bfseries B646} (2002) 220}
  [\href{https://arxiv.org/abs/hep-ph/0207004}{{\ttfamily hep-ph/0207004}}].

\bibitem{Anastasiou:2002wq}
C.~Anastasiou and K.~Melnikov, \emph{{Pseudoscalar Higgs boson production at
  hadron colliders in NNLO QCD}},
  \href{https://doi.org/10.1103/PhysRevD.67.037501}{\emph{Phys. Rev.}
  {\bfseries D67} (2003) 037501}
  [\href{https://arxiv.org/abs/hep-ph/0208115}{{\ttfamily hep-ph/0208115}}].

\bibitem{Ravindran:2003um}
V.~Ravindran, J.~Smith and W.~L. van Neerven, \emph{{NNLO corrections to the
  total cross-section for Higgs boson production in hadron hadron collisions}},
  \href{https://doi.org/10.1016/S0550-3213(03)00457-7}{\emph{Nucl. Phys.}
  {\bfseries B665} (2003) 325}
  [\href{https://arxiv.org/abs/hep-ph/0302135}{{\ttfamily hep-ph/0302135}}].

\bibitem{Harlander:2003ai}
R.~V. Harlander and W.~B. Kilgore, \emph{{Higgs boson production in bottom
  quark fusion at next-to-next-to leading order}},
  \href{https://doi.org/10.1103/PhysRevD.68.013001}{\emph{Phys. Rev.}
  {\bfseries D68} (2003) 013001}
  [\href{https://arxiv.org/abs/hep-ph/0304035}{{\ttfamily hep-ph/0304035}}].

\bibitem{Hamberg:1990np}
R.~Hamberg, W.~L. van Neerven and T.~Matsuura, \emph{{A complete calculation of
  the order $\alpha-s^{2}$ correction to the Drell-Yan $K$ factor}},
  \href{https://doi.org/10.1016/S0550-3213(02)00814-3,
  10.1016/0550-3213(91)90064-5}{\emph{Nucl. Phys.} {\bfseries B359} (1991)
  343}.

\bibitem{Anastasiou:2015ema}
C.~Anastasiou, C.~Duhr, F.~Dulat, F.~Herzog and B.~Mistlberger, \emph{{Higgs
  Boson Gluon-Fusion Production in QCD at Three Loops}},
  \href{https://doi.org/10.1103/PhysRevLett.114.212001}{\emph{Phys. Rev. Lett.}
  {\bfseries 114} (2015) 212001}
  [\href{https://arxiv.org/abs/1503.06056}{{\ttfamily 1503.06056}}].

\bibitem{Mistlberger:2018etf}
B.~Mistlberger, \emph{{Higgs boson production at hadron colliders at N$^{3}$LO
  in QCD}}, \href{https://doi.org/10.1007/JHEP05(2018)028}{\emph{JHEP}
  {\bfseries 05} (2018) 028}
  [\href{https://arxiv.org/abs/1802.00833}{{\ttfamily 1802.00833}}].

\bibitem{Duhr:2019kwi}
C.~Duhr, F.~Dulat and B.~Mistlberger, \emph{{Higgs production in bottom-quark
  fusion to third order in the strong coupling}},
  \href{https://arxiv.org/abs/1904.09990}{{\ttfamily 1904.09990}}.

\bibitem{Duhr:2020seh}
C.~Duhr, F.~Dulat and B.~Mistlberger, \emph{{The Drell-Yan cross section to
  third order in the strong coupling constant}},
  \href{https://arxiv.org/abs/2001.07717}{{\ttfamily 2001.07717}}.

\bibitem{Anastasiou:2004xq}
C.~Anastasiou, K.~Melnikov and F.~Petriello, \emph{{Higgs boson production at
  hadron colliders: Differential cross sections through next-to-next-to-leading
  order}}, \href{https://doi.org/10.1103/PhysRevLett.93.262002}{\emph{Phys.
  Rev. Lett.} {\bfseries 93} (2004) 262002}
  [\href{https://arxiv.org/abs/hep-ph/0409088}{{\ttfamily hep-ph/0409088}}].

\bibitem{Anastasiou:2011qx}
C.~Anastasiou, F.~Herzog and A.~Lazopoulos, \emph{{The fully differential decay
  rate of a Higgs boson to bottom-quarks at NNLO in QCD}},
  \href{https://doi.org/10.1007/JHEP03(2012)035}{\emph{JHEP} {\bfseries 03}
  (2012) 035} [\href{https://arxiv.org/abs/1110.2368}{{\ttfamily 1110.2368}}].

\bibitem{Buehler:2012cu}
S.~B{\"u}hler, F.~Herzog, A.~Lazopoulos and R.~M{\"u}ller, \emph{{The fully
  differential hadronic production of a Higgs boson via bottom quark fusion at
  NNLO}}, \href{https://doi.org/10.1007/JHEP07(2012)115}{\emph{JHEP} {\bfseries
  07} (2012) 115} [\href{https://arxiv.org/abs/1204.4415}{{\ttfamily
  1204.4415}}].

\bibitem{Dulat:2018bfe}
F.~Dulat, B.~Mistlberger and A.~Pelloni, \emph{{Precision predictions at
  N$^3$LO for the Higgs boson rapidity distribution at the LHC}},
  \href{https://doi.org/10.1103/PhysRevD.99.034004}{\emph{Phys. Rev.}
  {\bfseries D99} (2019) 034004}
  [\href{https://arxiv.org/abs/1810.09462}{{\ttfamily 1810.09462}}].

\bibitem{Cieri:2018oms}
L.~Cieri, X.~Chen, T.~Gehrmann, E.~W.~N. Glover and A.~Huss, \emph{{Higgs boson
  production at the LHC using the $q_T$ subtraction formalism at N$^3$LO QCD}},
  \href{https://doi.org/10.1007/JHEP02(2019)096}{\emph{JHEP} {\bfseries 02}
  (2019) 096} [\href{https://arxiv.org/abs/1807.11501}{{\ttfamily
  1807.11501}}].

\bibitem{Anastasiou:2003yy}
C.~Anastasiou, L.~J. Dixon, K.~Melnikov and F.~Petriello, \emph{{Dilepton
  rapidity distribution in the Drell-Yan process at NNLO in QCD}},
  \href{https://doi.org/10.1103/PhysRevLett.91.182002}{\emph{Phys. Rev. Lett.}
  {\bfseries 91} (2003) 182002}
  [\href{https://arxiv.org/abs/hep-ph/0306192}{{\ttfamily hep-ph/0306192}}].

\bibitem{Anastasiou:2003ds}
C.~Anastasiou, L.~J. Dixon, K.~Melnikov and F.~Petriello, \emph{{High precision
  QCD at hadron colliders: Electroweak gauge boson rapidity distributions at
  NNLO}}, \href{https://doi.org/10.1103/PhysRevD.69.094008}{\emph{Phys. Rev.}
  {\bfseries D69} (2004) 094008}
  [\href{https://arxiv.org/abs/hep-ph/0312266}{{\ttfamily hep-ph/0312266}}].

\bibitem{Catani:2009sm}
S.~Catani, L.~Cieri, G.~Ferrera, D.~de~Florian and M.~Grazzini, \emph{{Vector
  boson production at hadron colliders: a fully exclusive QCD calculation at
  NNLO}}, \href{https://doi.org/10.1103/PhysRevLett.103.082001}{\emph{Phys.
  Rev. Lett.} {\bfseries 103} (2009) 082001}
  [\href{https://arxiv.org/abs/0903.2120}{{\ttfamily 0903.2120}}].

\bibitem{Melnikov:2006kv}
K.~Melnikov and F.~Petriello, \emph{{Electroweak gauge boson production at
  hadron colliders through ${\cal O}(\alpha_s^2)$}},
  \href{https://doi.org/10.1103/PhysRevD.74.114017}{\emph{Phys. Rev.}
  {\bfseries D74} (2006) 114017}
  [\href{https://arxiv.org/abs/hep-ph/0609070}{{\ttfamily hep-ph/0609070}}].

\bibitem{Gavin:2012sy}
R.~Gavin, Y.~Li, F.~Petriello and S.~Quackenbush, \emph{{W Physics at the LHC
  with FEWZ 2.1}},
  \href{https://doi.org/10.1016/j.cpc.2012.09.005}{\emph{Comput. Phys. Commun.}
  {\bfseries 184} (2013) 208}
  [\href{https://arxiv.org/abs/1201.5896}{{\ttfamily 1201.5896}}].

\bibitem{Anastasiou:2014vaa}
C.~Anastasiou, C.~Duhr, F.~Dulat, E.~Furlan, T.~Gehrmann, F.~Herzog et~al.,
  \emph{{Higgs boson gluon fusion production at threshold in N$^3$LO QCD}},
  \href{https://doi.org/10.1016/j.physletb.2014.08.067}{\emph{Phys. Lett.}
  {\bfseries B737} (2014) 325}
  [\href{https://arxiv.org/abs/1403.4616}{{\ttfamily 1403.4616}}].

\bibitem{Moch:2005ky}
S.~Moch and A.~Vogt, \emph{{Higher-order soft corrections to lepton pair and
  Higgs boson production}},
  \href{https://doi.org/10.1016/j.physletb.2005.09.061}{\emph{Phys. Lett.}
  {\bfseries B631} (2005) 48}
  [\href{https://arxiv.org/abs/hep-ph/0508265}{{\ttfamily hep-ph/0508265}}].

\bibitem{Laenen:2005uz}
E.~Laenen and L.~Magnea, \emph{{Threshold resummation for electroweak
  annihilation from DIS data}},
  \href{https://doi.org/10.1016/j.physletb.2005.10.038}{\emph{Phys. Lett.}
  {\bfseries B632} (2006) 270}
  [\href{https://arxiv.org/abs/hep-ph/0508284}{{\ttfamily hep-ph/0508284}}].

\bibitem{Ravindran:2005vv}
V.~Ravindran, \emph{{On Sudakov and soft resummations in QCD}},
  \href{https://doi.org/10.1016/j.nuclphysb.2006.04.008}{\emph{Nucl. Phys.}
  {\bfseries B746} (2006) 58}
  [\href{https://arxiv.org/abs/hep-ph/0512249}{{\ttfamily hep-ph/0512249}}].

\bibitem{Ravindran:2006cg}
V.~Ravindran, \emph{{Higher-order threshold effects to inclusive processes in
  QCD}}, \href{https://doi.org/10.1016/j.nuclphysb.2006.06.025}{\emph{Nucl.
  Phys.} {\bfseries B752} (2006) 173}
  [\href{https://arxiv.org/abs/hep-ph/0603041}{{\ttfamily hep-ph/0603041}}].

\bibitem{Idilbi:2005ni}
A.~Idilbi, X.-d. Ji, J.-P. Ma and F.~Yuan, \emph{{Threshold resummation for
  Higgs production in effective field theory}},
  \href{https://doi.org/10.1103/PhysRevD.73.077501}{\emph{Phys. Rev.}
  {\bfseries D73} (2006) 077501}
  [\href{https://arxiv.org/abs/hep-ph/0509294}{{\ttfamily hep-ph/0509294}}].

\bibitem{Li:2014afw}
Y.~Li, A.~von Manteuffel, R.~M. Schabinger and H.~X. Zhu, \emph{{Soft-virtual
  corrections to Higgs production at N$^3$LO}},
  \href{https://doi.org/10.1103/PhysRevD.91.036008}{\emph{Phys. Rev.}
  {\bfseries D91} (2015) 036008}
  [\href{https://arxiv.org/abs/1412.2771}{{\ttfamily 1412.2771}}].

\bibitem{Ahmed:2014cha}
T.~Ahmed, N.~Rana and V.~Ravindran, \emph{{Higgs boson production through $b
  \bar b$ annihilation at threshold in N$^3$LO QCD}},
  \href{https://doi.org/10.1007/JHEP10(2014)139}{\emph{JHEP} {\bfseries 10}
  (2014) 139} [\href{https://arxiv.org/abs/1408.0787}{{\ttfamily 1408.0787}}].

\bibitem{Ravindran:2006bu}
V.~Ravindran, J.~Smith and W.~L. van Neerven, \emph{{QCD threshold corrections
  to di-lepton and Higgs rapidity distributions beyond $N^{2}$ LO}},
  \href{https://doi.org/10.1016/j.nuclphysb.2007.01.005}{\emph{Nucl. Phys.}
  {\bfseries B767} (2007) 100}
  [\href{https://arxiv.org/abs/hep-ph/0608308}{{\ttfamily hep-ph/0608308}}].

\bibitem{Ahmed:2014cla}
T.~Ahmed, M.~Mahakhud, N.~Rana and V.~Ravindran, \emph{{Drell-Yan Production at
  Threshold to Third Order in QCD}},
  \href{https://doi.org/10.1103/PhysRevLett.113.112002}{\emph{Phys. Rev. Lett.}
  {\bfseries 113} (2014) 112002}
  [\href{https://arxiv.org/abs/1404.0366}{{\ttfamily 1404.0366}}].

\bibitem{Kumar:2014uwa}
M.~C. Kumar, M.~K. Mandal and V.~Ravindran, \emph{{Associated production of
  Higgs boson with vector boson at threshold N$^{3}$LO in QCD}},
  \href{https://doi.org/10.1007/JHEP03(2015)037}{\emph{JHEP} {\bfseries 03}
  (2015) 037} [\href{https://arxiv.org/abs/1412.3357}{{\ttfamily 1412.3357}}].

\bibitem{Ahmed:2015qda}
T.~Ahmed, M.~C. Kumar, P.~Mathews, N.~Rana and V.~Ravindran,
  \emph{{Pseudo-scalar Higgs boson production at threshold N$^3$ LO and N$^3$
  LL QCD}}, \href{https://doi.org/10.1140/epjc/s10052-016-4199-1}{\emph{Eur.
  Phys. J.} {\bfseries C76} (2016) 355}
  [\href{https://arxiv.org/abs/1510.02235}{{\ttfamily 1510.02235}}].

\bibitem{Ravindran:2007sv}
V.~Ravindran and J.~Smith, \emph{{Threshold corrections to rapidity
  distributions of $Z$ and $W^\pm$ bosons beyond $N^{2}$ LO at hadron
  colliders}}, \href{https://doi.org/10.1103/PhysRevD.76.114004}{\emph{Phys.
  Rev.} {\bfseries D76} (2007) 114004}
  [\href{https://arxiv.org/abs/0708.1689}{{\ttfamily 0708.1689}}].

\bibitem{Ahmed:2014uya}
T.~Ahmed, M.~K. Mandal, N.~Rana and V.~Ravindran, \emph{{Rapidity Distributions
  in Drell-Yan and Higgs Productions at Threshold to Third Order in QCD}},
  \href{https://doi.org/10.1103/PhysRevLett.113.212003}{\emph{Phys. Rev. Lett.}
  {\bfseries 113} (2014) 212003}
  [\href{https://arxiv.org/abs/1404.6504}{{\ttfamily 1404.6504}}].

\bibitem{Ahmed:2014era}
T.~Ahmed, M.~K. Mandal, N.~Rana and V.~Ravindran, \emph{{Higgs Rapidity
  Distribution in $b {\bar b}$ Annihilation at Threshold in N$^{3}$LO QCD}},
  \href{https://doi.org/10.1007/JHEP02(2015)131}{\emph{JHEP} {\bfseries 02}
  (2015) 131} [\href{https://arxiv.org/abs/1411.5301}{{\ttfamily 1411.5301}}].

\bibitem{Catani:2003zt}
S.~Catani, D.~de~Florian, M.~Grazzini and P.~Nason, \emph{{Soft gluon
  resummation for Higgs boson production at hadron colliders}},
  \href{https://doi.org/10.1088/1126-6708/2003/07/028}{\emph{JHEP} {\bfseries
  07} (2003) 028} [\href{https://arxiv.org/abs/hep-ph/0306211}{{\ttfamily
  hep-ph/0306211}}].

\bibitem{Catani:2014uta}
S.~Catani, L.~Cieri, D.~de~Florian, G.~Ferrera and M.~Grazzini,
  \emph{{Threshold resummation at N$^3$LL accuracy and soft-virtual cross
  sections at N$^3$LO}},
  \href{https://doi.org/10.1016/j.nuclphysb.2014.09.012}{\emph{Nucl. Phys.}
  {\bfseries B888} (2014) 75}
  [\href{https://arxiv.org/abs/1405.4827}{{\ttfamily 1405.4827}}].

\bibitem{Bonvini:2014joa}
M.~Bonvini and S.~Marzani, \emph{{Resummed Higgs cross section at N$^{3}$LL}},
  \href{https://doi.org/10.1007/JHEP09(2014)007}{\emph{JHEP} {\bfseries 09}
  (2014) 007} [\href{https://arxiv.org/abs/1405.3654}{{\ttfamily 1405.3654}}].

\bibitem{Ahmed:2015sna}
T.~Ahmed, G.~Das, M.~C. Kumar, N.~Rana and V.~Ravindran, \emph{{RG improved
  Higgs boson production to N$^3$LO in QCD}},
  \href{https://arxiv.org/abs/1505.07422}{{\ttfamily 1505.07422}}.

\bibitem{Bonvini:2016frm}
M.~Bonvini, S.~Marzani, C.~Muselli and L.~Rottoli, \emph{{On the Higgs cross
  section at N$^{3}$LO+N$^{3}$LL and its uncertainty}},
  \href{https://doi.org/10.1007/JHEP08(2016)105}{\emph{JHEP} {\bfseries 08}
  (2016) 105} [\href{https://arxiv.org/abs/1603.08000}{{\ttfamily
  1603.08000}}].

\bibitem{H:2019dcl}
A.~A~H, A.~Chakraborty, G.~Das, P.~Mukherjee and V.~Ravindran, \emph{{Resummed
  prediction for Higgs boson production through $b\bar{b}$ annihilation at
  N$^3$LO+N$^3$LL}},  \href{https://arxiv.org/abs/1905.03771}{{\ttfamily
  1905.03771}}.

\bibitem{Moch:2005ba}
S.~Moch, J.~A.~M. Vermaseren and A.~Vogt, \emph{{Higher-order corrections in
  threshold resummation}},
  \href{https://doi.org/10.1016/j.nuclphysb.2005.08.005}{\emph{Nucl. Phys.}
  {\bfseries B726} (2005) 317}
  [\href{https://arxiv.org/abs/hep-ph/0506288}{{\ttfamily hep-ph/0506288}}].

\bibitem{Schmidt:2015cea}
T.~Schmidt and M.~Spira, \emph{{Higgs Boson Production via Gluon Fusion:
  Soft-Gluon Resummation including Mass Effects}},
  \href{https://doi.org/10.1103/PhysRevD.93.014022}{\emph{Phys. Rev.}
  {\bfseries D93} (2016) 014022}
  [\href{https://arxiv.org/abs/1509.00195}{{\ttfamily 1509.00195}}].

\bibitem{deFlorian:2007sr}
D.~de~Florian and J.~Zurita, \emph{{Soft-gluon resummation for pseudoscalar
  Higgs boson production at hadron colliders}},
  \href{https://doi.org/10.1016/j.physletb.2007.11.018}{\emph{Phys. Lett.}
  {\bfseries B659} (2008) 813}
  [\href{https://arxiv.org/abs/0711.1916}{{\ttfamily 0711.1916}}].

\bibitem{Ahmed:2016otz}
T.~Ahmed, M.~Bonvini, M.~C. Kumar, P.~Mathews, N.~Rana, V.~Ravindran et~al.,
  \emph{{Pseudo-scalar Higgs boson production at N$^3$ LO$_{\text {A}}$ +N$^3$
  LL $'$}}, \href{https://doi.org/10.1140/epjc/s10052-016-4510-1}{\emph{Eur.
  Phys. J.} {\bfseries C76} (2016) 663}
  [\href{https://arxiv.org/abs/1606.00837}{{\ttfamily 1606.00837}}].

\bibitem{Das:2019btv}
G.~Das, S.-O. Moch and A.~Vogt, \emph{{Soft corrections to inclusive
  deep-inelastic scattering at four loops and beyond}},
  \href{https://arxiv.org/abs/1912.12920}{{\ttfamily 1912.12920}}.

\bibitem{Catani:1989ne}
S.~Catani and L.~Trentadue, \emph{{Resummation of the QCD Perturbative Series
  for Hard Processes}},
  \href{https://doi.org/10.1016/0550-3213(89)90273-3}{\emph{Nucl. Phys.}
  {\bfseries B327} (1989) 323}.

\bibitem{Westmark:2017uig}
D.~Westmark and J.~F. Owens, \emph{{Enhanced threshold resummation formalism
  for lepton pair production and its effects in the determination of parton
  distribution functions}},
  \href{https://doi.org/10.1103/PhysRevD.95.056024}{\emph{Phys. Rev.}
  {\bfseries D95} (2017) 056024}
  [\href{https://arxiv.org/abs/1701.06716}{{\ttfamily 1701.06716}}].

\bibitem{Banerjee:2017cfc}
P.~Banerjee, G.~Das, P.~K. Dhani and V.~Ravindran, \emph{{Threshold resummation
  of the rapidity distribution for Higgs production at NNLO+NNLL}},
  \href{https://doi.org/10.1103/PhysRevD.97.054024}{\emph{Phys. Rev.}
  {\bfseries D97} (2018) 054024}
  [\href{https://arxiv.org/abs/1708.05706}{{\ttfamily 1708.05706}}].

\bibitem{Banerjee:2018vvb}
P.~Banerjee, G.~Das, P.~K. Dhani and V.~Ravindran, \emph{{Threshold resummation
  of the rapidity distribution for Drell-Yan production at NNLO+NNLL}},
  \href{https://doi.org/10.1103/PhysRevD.98.054018}{\emph{Phys. Rev.}
  {\bfseries D98} (2018) 054018}
  [\href{https://arxiv.org/abs/1805.01186}{{\ttfamily 1805.01186}}].

\bibitem{Lustermans:2019cau}
G.~Lustermans, J.~K.~L. Michel and F.~J. Tackmann, \emph{{Generalized Threshold
  Factorization with Full Collinear Dynamics}},
  \href{https://arxiv.org/abs/1908.00985}{{\ttfamily 1908.00985}}.

\bibitem{Ebert:2017uel}
M.~A. Ebert, J.~K.~L. Michel and F.~J. Tackmann, \emph{{Resummation Improved
  Rapidity Spectrum for Gluon Fusion Higgs Production}},
  \href{https://doi.org/10.1007/JHEP05(2017)088}{\emph{JHEP} {\bfseries 05}
  (2017) 088} [\href{https://arxiv.org/abs/1702.00794}{{\ttfamily
  1702.00794}}].

\bibitem{Mathews:2004xp}
P.~Mathews, V.~Ravindran, K.~Sridhar and W.~L. van Neerven,
  \emph{{Next-to-leading order QCD corrections to the Drell-Yan cross section
  in models of TeV-scale gravity}},
  \href{https://doi.org/10.1016/j.nuclphysb.2005.01.051}{\emph{Nucl. Phys.}
  {\bfseries B713} (2005) 333}
  [\href{https://arxiv.org/abs/hep-ph/0411018}{{\ttfamily hep-ph/0411018}}].

\bibitem{Kumar:2007af}
M.~C. Kumar, P.~Mathews, V.~Ravindran and A.~Tripathi, \emph{{Unparticle
  physics in diphoton production at the CERN LHC}},
  \href{https://doi.org/10.1103/PhysRevD.77.055013}{\emph{Phys. Rev.}
  {\bfseries D77} (2008) 055013}
  [\href{https://arxiv.org/abs/0709.2478}{{\ttfamily 0709.2478}}].

\bibitem{Kumar:2008dn}
M.~C. Kumar, P.~Mathews, V.~Ravindran and A.~Tripathi, \emph{{Unparticles in
  diphoton production to NLO in QCD at the LHC}},
  \href{https://doi.org/10.1103/PhysRevD.79.075012}{\emph{Phys. Rev.}
  {\bfseries D79} (2009) 075012}
  [\href{https://arxiv.org/abs/0804.4054}{{\ttfamily 0804.4054}}].

\bibitem{Kumar:2008pk}
M.~C. Kumar, P.~Mathews, V.~Ravindran and A.~Tripathi, \emph{{Diphoton signals
  in theories with large extra dimensions to NLO QCD at hadron colliders}},
  \href{https://doi.org/10.1016/j.physletb.2009.01.002}{\emph{Phys. Lett.}
  {\bfseries B672} (2009) 45}
  [\href{https://arxiv.org/abs/0811.1670}{{\ttfamily 0811.1670}}].

\bibitem{Kumar:2009nn}
M.~C. Kumar, P.~Mathews, V.~Ravindran and A.~Tripathi, \emph{{Direct photon
  pair production at the LHC to order $\alpha_s$ in TeV scale gravity models}},
  \href{https://doi.org/10.1016/j.nuclphysb.2009.03.022}{\emph{Nucl. Phys.}
  {\bfseries B818} (2009) 28}
  [\href{https://arxiv.org/abs/0902.4894}{{\ttfamily 0902.4894}}].

\bibitem{Agarwal:2009xr}
N.~Agarwal, V.~Ravindran, V.~K. Tiwari and A.~Tripathi, \emph{{Z boson pair
  production at the LHC to ${\cal O}(\alpha_s)$ in TeV scale gravity models}},
  \href{https://doi.org/10.1016/j.nuclphysb.2009.12.032}{\emph{Nucl. Phys.}
  {\bfseries B830} (2010) 248}
  [\href{https://arxiv.org/abs/0909.2651}{{\ttfamily 0909.2651}}].

\bibitem{Agarwal:2010sp}
N.~Agarwal, V.~Ravindran, V.~K. Tiwari and A.~Tripathi, \emph{{$W^+W^-$
  production in Large extra dimension model at next-to-leading order in QCD at
  the LHC}}, \href{https://doi.org/10.1103/PhysRevD.82.036001}{\emph{Phys.
  Rev.} {\bfseries D82} (2010) 036001}
  [\href{https://arxiv.org/abs/1003.5450}{{\ttfamily 1003.5450}}].

\bibitem{Kumar:2011yta}
M.~C. Kumar, P.~Mathews, A.~A. Pankov, N.~Paver, V.~Ravindran and A.~V.
  Tsytrinov, \emph{{Spin-analysis of s-channel diphoton resonances at the
  LHC}}, \href{https://doi.org/10.1103/PhysRevD.84.115008,
  10.1103/PhysRevD.84.119902}{\emph{Phys. Rev.} {\bfseries D84} (2011) 115008}
  [\href{https://arxiv.org/abs/1108.3764}{{\ttfamily 1108.3764}}].

\bibitem{Agarwal:2009zg}
N.~Agarwal, V.~Ravindran, V.~K. Tiwari and A.~Tripathi, \emph{{Next-to-leading
  order QCD corrections to the $Z$ boson pair production at the LHC in Randall
  Sundrum model}},
  \href{https://doi.org/10.1016/j.physletb.2010.02.060}{\emph{Phys. Lett.}
  {\bfseries B686} (2010) 244}
  [\href{https://arxiv.org/abs/0910.1551}{{\ttfamily 0910.1551}}].

\bibitem{Agarwal:2010sn}
N.~Agarwal, V.~Ravindran, V.~K. Tiwari and A.~Tripathi, \emph{{Next-to-leading
  order QCD corrections to $W^+W^-$ production at the LHC in Randall Sundrum
  model}}, \href{https://doi.org/10.1016/j.physletb.2010.05.063}{\emph{Phys.
  Lett.} {\bfseries B690} (2010) 390}
  [\href{https://arxiv.org/abs/1003.5445}{{\ttfamily 1003.5445}}].

\bibitem{Mathews:2005bw}
P.~Mathews, V.~Ravindran and K.~Sridhar, \emph{{NLO-QCD corrections to dilepton
  production in the Randall-Sundrum model}},
  \href{https://doi.org/10.1088/1126-6708/2005/10/031}{\emph{JHEP} {\bfseries
  10} (2005) 031} [\href{https://arxiv.org/abs/hep-ph/0506158}{{\ttfamily
  hep-ph/0506158}}].

\bibitem{ArkaniHamed:1998rs}
N.~Arkani-Hamed, S.~Dimopoulos and G.~R. Dvali, \emph{{The Hierarchy problem
  and new dimensions at a millimeter}},
  \href{https://doi.org/10.1016/S0370-2693(98)00466-3}{\emph{Phys. Lett.}
  {\bfseries B429} (1998) 263}
  [\href{https://arxiv.org/abs/hep-ph/9803315}{{\ttfamily hep-ph/9803315}}].

\bibitem{Frederix:2012dp}
R.~Frederix, M.~K. Mandal, P.~Mathews, V.~Ravindran, S.~Seth, P.~Torrielli
  et~al., \emph{{Diphoton production in the ADD model to NLO+parton shower
  accuracy at the LHC}},
  \href{https://doi.org/10.1007/JHEP12(2012)102}{\emph{JHEP} {\bfseries 12}
  (2012) 102} [\href{https://arxiv.org/abs/1209.6527}{{\ttfamily 1209.6527}}].

\bibitem{Frederix:2013lga}
R.~Frederix, M.~K. Mandal, P.~Mathews, V.~Ravindran and S.~Seth,
  \emph{{Drell-Yan, $ZZ, W^+W^-$ production in SM \& ADD model to NLO+PS
  accuracy at the LHC}},
  \href{https://doi.org/10.1140/epjc/s10052-014-2745-2}{\emph{Eur. Phys. J.}
  {\bfseries C74} (2014) 2745}
  [\href{https://arxiv.org/abs/1307.7013}{{\ttfamily 1307.7013}}].

\bibitem{Das:2014tva}
G.~Das, P.~Mathews, V.~Ravindran and S.~Seth, \emph{{RS resonance in di-final
  state production at the LHC to NLO+PS accuracy}},
  \href{https://doi.org/10.1007/JHEP10(2014)188}{\emph{JHEP} {\bfseries 10}
  (2014) 188} [\href{https://arxiv.org/abs/1408.3970}{{\ttfamily 1408.3970}}].

\bibitem{Kumar:2010kv}
M.~C. Kumar, P.~Mathews, V.~Ravindran and S.~Seth, \emph{{Vector boson
  production in association with KK modes of the ADD model to NLO in QCD at
  LHC}}, \href{https://doi.org/10.1088/0954-3899/38/5/055001}{\emph{J. Phys.}
  {\bfseries G38} (2011) 055001}
  [\href{https://arxiv.org/abs/1004.5519}{{\ttfamily 1004.5519}}].

\bibitem{Kumar:2011jq}
M.~C. Kumar, P.~Mathews, V.~Ravindran and S.~Seth, \emph{{Neutral triple
  electroweak gauge boson production in the large extra-dimension model at the
  LHC}}, \href{https://doi.org/10.1103/PhysRevD.85.094507}{\emph{Phys. Rev.}
  {\bfseries D85} (2012) 094507}
  [\href{https://arxiv.org/abs/1111.7063}{{\ttfamily 1111.7063}}].

\bibitem{Das:2015bda}
G.~Das and P.~Mathews, \emph{{Neutral Triple Vector Boson Production in
  Randall-Sundrum Model at the LHC}},
  \href{https://doi.org/10.1103/PhysRevD.92.094034}{\emph{Phys. Rev.}
  {\bfseries D92} (2015) 094034}
  [\href{https://arxiv.org/abs/1507.08857}{{\ttfamily 1507.08857}}].

\bibitem{Das:2016pbk}
G.~Das, C.~Degrande, V.~Hirschi, F.~Maltoni and H.-S. Shao, \emph{{NLO
  predictions for the production of a spin-two particle at the LHC}},
  \href{https://doi.org/10.1016/j.physletb.2017.05.007}{\emph{Phys. Lett.}
  {\bfseries B770} (2017) 507}
  [\href{https://arxiv.org/abs/1605.09359}{{\ttfamily 1605.09359}}].

\bibitem{Alloul:2013bka}
A.~Alloul, N.~D. Christensen, C.~Degrande, C.~Duhr and B.~Fuks,
  \emph{{FeynRules 2.0 - A complete toolbox for tree-level phenomenology}},
  \href{https://doi.org/10.1016/j.cpc.2014.04.012}{\emph{Comput. Phys. Commun.}
  {\bfseries 185} (2014) 2250}
  [\href{https://arxiv.org/abs/1310.1921}{{\ttfamily 1310.1921}}].

\bibitem{Alwall:2014hca}
J.~Alwall, R.~Frederix, S.~Frixione, V.~Hirschi, F.~Maltoni, O.~Mattelaer
  et~al., \emph{{The automated computation of tree-level and next-to-leading
  order differential cross sections, and their matching to parton shower
  simulations}}, \href{https://doi.org/10.1007/JHEP07(2014)079}{\emph{JHEP}
  {\bfseries 07} (2014) 079} [\href{https://arxiv.org/abs/1405.0301}{{\ttfamily
  1405.0301}}].

\bibitem{deFlorian:2013wpa}
D.~de~Florian, M.~Mahakhud, P.~Mathews, J.~Mazzitelli and V.~Ravindran,
  \emph{{Next-to-Next-to-Leading Order QCD Corrections in Models of TeV-Scale
  Gravity}}, \href{https://doi.org/10.1007/JHEP04(2014)028}{\emph{JHEP}
  {\bfseries 04} (2014) 028} [\href{https://arxiv.org/abs/1312.7173}{{\ttfamily
  1312.7173}}].

\bibitem{deFlorian:2013sza}
D.~de~Florian, M.~Mahakhud, P.~Mathews, J.~Mazzitelli and V.~Ravindran,
  \emph{{Quark and gluon spin-2 form factors to two-loops in QCD}},
  \href{https://doi.org/10.1007/JHEP02(2014)035}{\emph{JHEP} {\bfseries 02}
  (2014) 035} [\href{https://arxiv.org/abs/1312.6528}{{\ttfamily 1312.6528}}].

\bibitem{Ahmed:2016qhu}
T.~Ahmed, P.~Banerjee, P.~K. Dhani, M.~C. Kumar, P.~Mathews, N.~Rana et~al.,
  \emph{{NNLO QCD corrections to the Drell-Yan cross section in models of
  TeV-scale gravity}},
  \href{https://doi.org/10.1140/epjc/s10052-016-4587-6}{\emph{Eur. Phys. J.}
  {\bfseries C77} (2017) 22}
  [\href{https://arxiv.org/abs/1606.08454}{{\ttfamily 1606.08454}}].

\bibitem{Banerjee:2017ewt}
P.~Banerjee, P.~K. Dhani, M.~C. Kumar, P.~Mathews and V.~Ravindran, \emph{{NNLO
  QCD corrections to production of a spin-2 particle with nonuniversal
  couplings in the Drell-Yan process}},
  \href{https://doi.org/10.1103/PhysRevD.97.094028}{\emph{Phys. Rev.}
  {\bfseries D97} (2018) 094028}
  [\href{https://arxiv.org/abs/1710.04184}{{\ttfamily 1710.04184}}].

\bibitem{Das:2019bxi}
G.~Das, M.~C. Kumar and K.~Samanta, \emph{{Resummed inclusive cross-section in
  ADD model at N$^3$LL+NNLO}},
  \href{https://arxiv.org/abs/1912.13039}{{\ttfamily 1912.13039}}.

\bibitem{Ahmed:2015qia}
T.~Ahmed, G.~Das, P.~Mathews, N.~Rana and V.~Ravindran, \emph{{Spin-2 Form
  Factors at Three Loop in QCD}},
  \href{https://doi.org/10.1007/JHEP12(2015)084}{\emph{JHEP} {\bfseries 12}
  (2015) 084} [\href{https://arxiv.org/abs/1508.05043}{{\ttfamily
  1508.05043}}].

\bibitem{Davoudiasl:1999jd}
H.~Davoudiasl, J.~L. Hewett and T.~G. Rizzo, \emph{{Phenomenology of the
  Randall-Sundrum Gauge Hierarchy Model}},
  \href{https://doi.org/10.1103/PhysRevLett.84.2080}{\emph{Phys. Rev. Lett.}
  {\bfseries 84} (2000) 2080}
  [\href{https://arxiv.org/abs/hep-ph/9909255}{{\ttfamily hep-ph/9909255}}].

\bibitem{Goldberger:1999un}
W.~D. Goldberger and M.~B. Wise, \emph{{Phenomenology of a stabilized
  modulus}}, \href{https://doi.org/10.1016/S0370-2693(00)00099-X}{\emph{Phys.
  Lett.} {\bfseries B475} (2000) 275}
  [\href{https://arxiv.org/abs/hep-ph/9911457}{{\ttfamily hep-ph/9911457}}].

\bibitem{Goldberger:1999uk}
W.~D. Goldberger and M.~B. Wise, \emph{{Modulus stabilization with bulk
  fields}}, \href{https://doi.org/10.1103/PhysRevLett.83.4922}{\emph{Phys. Rev.
  Lett.} {\bfseries 83} (1999) 4922}
  [\href{https://arxiv.org/abs/hep-ph/9907447}{{\ttfamily hep-ph/9907447}}].

\bibitem{Han:1998sg}
T.~Han, J.~D. Lykken and R.-J. Zhang, \emph{{On Kaluza-Klein states from large
  extra dimensions}},
  \href{https://doi.org/10.1103/PhysRevD.59.105006}{\emph{Phys. Rev.}
  {\bfseries D59} (1999) 105006}
  [\href{https://arxiv.org/abs/hep-ph/9811350}{{\ttfamily hep-ph/9811350}}].

\bibitem{Giudice:1998ck}
G.~F. Giudice, R.~Rattazzi and J.~D. Wells, \emph{{Quantum gravity and extra
  dimensions at high-energy colliders}},
  \href{https://doi.org/10.1016/S0550-3213(99)00044-9}{\emph{Nucl. Phys.}
  {\bfseries B544} (1999) 3}
  [\href{https://arxiv.org/abs/hep-ph/9811291}{{\ttfamily hep-ph/9811291}}].

\bibitem{Davoudiasl:2000wi}
H.~Davoudiasl, J.~L. Hewett and T.~G. Rizzo, \emph{{Experimental probes of
  localized gravity: On and off the wall}},
  \href{https://doi.org/10.1103/PhysRevD.63.075004}{\emph{Phys. Rev.}
  {\bfseries D63} (2001) 075004}
  [\href{https://arxiv.org/abs/hep-ph/0006041}{{\ttfamily hep-ph/0006041}}].

\bibitem{KumarRai:2003kk}
S.~Kumar~Rai and S.~Raychaudhuri, \emph{{Single photon signals for warped
  quantum gravity at a linear e+ e- collider}},
  \href{https://doi.org/10.1088/1126-6708/2003/10/020}{\emph{JHEP} {\bfseries
  10} (2003) 020} [\href{https://arxiv.org/abs/hep-ph/0307096}{{\ttfamily
  hep-ph/0307096}}].

\bibitem{Altarelli:1978id}
G.~Altarelli, R.~K. Ellis and G.~Martinelli, \emph{{Leptoproduction and
  Drell-Yan Processes Beyond the Leading Approximation in Chromodynamics}},
  \href{https://doi.org/10.1016/0550-3213(78)90085-8,
  10.1016/0550-3213(78)90067-6}{\emph{Nucl. Phys.} {\bfseries B143} (1978)
  521}.

\bibitem{Matsuura:1987wt}
T.~Matsuura and W.~L. van Neerven, \emph{{Second Order Logarithmic Corrections
  to the {Drell-Yan} Cross-section}},
  \href{https://doi.org/10.1007/BF01624369}{\emph{Z. Phys.} {\bfseries C38}
  (1988) 623}.

\bibitem{Matsuura:1988sm}
T.~Matsuura, S.~C. van~der Marck and W.~L. van Neerven, \emph{{The Calculation
  of the Second Order Soft and Virtual Contributions to the Drell-Yan
  Cross-Section}},
  \href{https://doi.org/10.1016/0550-3213(89)90620-2}{\emph{Nucl. Phys.}
  {\bfseries B319} (1989) 570}.

\bibitem{Catani:1996yz}
S.~Catani, M.~L. Mangano, P.~Nason and L.~Trentadue, \emph{{The Resummation of
  soft gluons in hadronic collisions}},
  \href{https://doi.org/10.1016/0550-3213(96)00399-9}{\emph{Nucl. Phys.}
  {\bfseries B478} (1996) 273}
  [\href{https://arxiv.org/abs/hep-ph/9604351}{{\ttfamily hep-ph/9604351}}].

\bibitem{Moch:2004pa}
S.~Moch, J.~A.~M. Vermaseren and A.~Vogt, \emph{{The Three loop splitting
  functions in QCD: The Nonsinglet case}},
  \href{https://doi.org/10.1016/j.nuclphysb.2004.03.030}{\emph{Nucl. Phys.}
  {\bfseries B688} (2004) 101}
  [\href{https://arxiv.org/abs/hep-ph/0403192}{{\ttfamily hep-ph/0403192}}].

\bibitem{Lee:2016ixa}
J.~Henn, A.~V. Smirnov, V.~A. Smirnov, M.~Steinhauser and R.~N. Lee,
  \emph{{Four-loop photon quark form factor and cusp anomalous dimension in the
  large-$N_c$ limit of QCD}},
  \href{https://doi.org/10.1007/JHEP03(2017)139}{\emph{JHEP} {\bfseries 03}
  (2017) 139} [\href{https://arxiv.org/abs/1612.04389}{{\ttfamily
  1612.04389}}].

\bibitem{Moch:2017uml}
S.~Moch, B.~Ruijl, T.~Ueda, J.~A.~M. Vermaseren and A.~Vogt, \emph{{Four-Loop
  Non-Singlet Splitting Functions in the Planar Limit and Beyond}},
  \href{https://doi.org/10.1007/JHEP10(2017)041}{\emph{JHEP} {\bfseries 10}
  (2017) 041} [\href{https://arxiv.org/abs/1707.08315}{{\ttfamily
  1707.08315}}].

\bibitem{Grozin:2018vdn}
A.~Grozin, \emph{{Four-loop cusp anomalous dimension in QED}},
  \href{https://doi.org/10.1007/JHEP06(2018)073,
  10.1007/JHEP01(2019)134}{\emph{JHEP} {\bfseries 06} (2018) 073}
  [\href{https://arxiv.org/abs/1805.05050}{{\ttfamily 1805.05050}}].

\bibitem{Henn:2019rmi}
J.~M. Henn, T.~Peraro, M.~Stahlhofen and P.~Wasser, \emph{{Matter dependence of
  the four-loop cusp anomalous dimension}},
  \href{https://doi.org/10.1103/PhysRevLett.122.201602}{\emph{Phys. Rev. Lett.}
  {\bfseries 122} (2019) 201602}
  [\href{https://arxiv.org/abs/1901.03693}{{\ttfamily 1901.03693}}].

\bibitem{Bruser:2019auj}
R.~Br{\"u}ser, A.~Grozin, J.~M. Henn and M.~Stahlhofen, \emph{{Matter
  dependence of the four-loop QCD cusp anomalous dimension: from small angles
  to all angles}}, \href{https://doi.org/10.1007/JHEP05(2019)186}{\emph{JHEP}
  {\bfseries 05} (2019) 186}
  [\href{https://arxiv.org/abs/1902.05076}{{\ttfamily 1902.05076}}].

\bibitem{Davies:2016jie}
J.~Davies, A.~Vogt, B.~Ruijl, T.~Ueda and J.~A.~M. Vermaseren,
  \emph{{Large-$n_f$ contributions to the four-loop splitting functions in
  QCD}}, \href{https://doi.org/10.1016/j.nuclphysb.2016.12.012}{\emph{Nucl.
  Phys.} {\bfseries B915} (2017) 335}
  [\href{https://arxiv.org/abs/1610.07477}{{\ttfamily 1610.07477}}].

\bibitem{Lee:2017mip}
R.~N. Lee, A.~V. Smirnov, V.~A. Smirnov and M.~Steinhauser, \emph{{The $n_f^2$
  contributions to fermionic four-loop form factors}},
  \href{https://doi.org/10.1103/PhysRevD.96.014008}{\emph{Phys. Rev.}
  {\bfseries D96} (2017) 014008}
  [\href{https://arxiv.org/abs/1705.06862}{{\ttfamily 1705.06862}}].

\bibitem{Gracey:1994nn}
J.~A. Gracey, \emph{{Anomalous dimension of nonsinglet Wilson operators at
  ${\cal O}(1/n_f)$ in deep inelastic scattering}},
  \href{https://doi.org/10.1016/0370-2693(94)90502-9}{\emph{Phys. Lett.}
  {\bfseries B322} (1994) 141}
  [\href{https://arxiv.org/abs/hep-ph/9401214}{{\ttfamily hep-ph/9401214}}].

\bibitem{Beneke:1995pq}
M.~Beneke and V.~M. Braun, \emph{{Power corrections and renormalons in
  Drell-Yan production}},
  \href{https://doi.org/10.1016/0550-3213(95)00439-Y}{\emph{Nucl. Phys.}
  {\bfseries B454} (1995) 253}
  [\href{https://arxiv.org/abs/hep-ph/9506452}{{\ttfamily hep-ph/9506452}}].

\bibitem{Moch:2018wjh}
S.~Moch, B.~Ruijl, T.~Ueda, J.~A.~M. Vermaseren and A.~Vogt, \emph{{On quartic
  colour factors in splitting functions and the gluon cusp anomalous
  dimension}},
  \href{https://doi.org/10.1016/j.physletb.2018.06.017}{\emph{Phys. Lett.}
  {\bfseries B782} (2018) 627}
  [\href{https://arxiv.org/abs/1805.09638}{{\ttfamily 1805.09638}}].

\bibitem{Lee:2019zop}
R.~N. Lee, A.~V. Smirnov, V.~A. Smirnov and M.~Steinhauser, \emph{{Four-loop
  quark form factor with quartic fundamental colour factor}},
  \href{https://doi.org/10.1007/JHEP02(2019)172}{\emph{JHEP} {\bfseries 02}
  (2019) 172} [\href{https://arxiv.org/abs/1901.02898}{{\ttfamily
  1901.02898}}].

\bibitem{Henn:2019swt}
J.~M. Henn, G.~P. Korchemsky and B.~Mistlberger, \emph{{The full four-loop cusp
  anomalous dimension in $\mathcal{N}=4$ super Yang-Mills and QCD}},
  \href{https://arxiv.org/abs/1911.10174}{{\ttfamily 1911.10174}}.

\bibitem{H.:2020ecd}
A.~A. H., G.~Das, M.~C. Kumar, P.~Mukherjee, V.~Ravindran and K.~Samanta,
  \emph{{Resummed Drell-Yan cross-section at N$^3$LL}},
  \href{https://arxiv.org/abs/2001.11377}{{\ttfamily 2001.11377}}.

\bibitem{Vogt:2004ns}
A.~Vogt, \emph{{Efficient evolution of unpolarized and polarized parton
  distributions with QCD-PEGASUS}},
  \href{https://doi.org/10.1016/j.cpc.2005.03.103}{\emph{Comput. Phys. Commun.}
  {\bfseries 170} (2005) 65}
  [\href{https://arxiv.org/abs/hep-ph/0408244}{{\ttfamily hep-ph/0408244}}].

\bibitem{Buckley:2014ana}
A.~Buckley, J.~Ferrando, S.~Lloyd, K.~Nordstr{\"o}m, B.~Page, M.~R{\"u}fenacht
  et~al., \emph{{LHAPDF6: parton density access in the LHC precision era}},
  \href{https://doi.org/10.1140/epjc/s10052-015-3318-8}{\emph{Eur. Phys. J.}
  {\bfseries C75} (2015) 132}
  [\href{https://arxiv.org/abs/1412.7420}{{\ttfamily 1412.7420}}].

\bibitem{Vermaseren:2000nd}
J.~A.~M. Vermaseren, \emph{{New features of FORM}},
  \href{https://arxiv.org/abs/math-ph/0010025}{{\ttfamily math-ph/0010025}}.

\bibitem{Ruijl:2017dtg}
B.~Ruijl, T.~Ueda and J.~Vermaseren, \emph{{FORM version 4.2}},
  \href{https://arxiv.org/abs/1707.06453}{{\ttfamily 1707.06453}}.

\bibitem{Harland-Lang:2014zoa}
L.~A. Harland-Lang, A.~D. Martin, P.~Motylinski and R.~S. Thorne, \emph{{Parton
  distributions in the LHC era: MMHT 2014 PDFs}},
  \href{https://doi.org/10.1140/epjc/s10052-015-3397-6}{\emph{Eur. Phys. J.}
  {\bfseries C75} (2015) 204}
  [\href{https://arxiv.org/abs/1412.3989}{{\ttfamily 1412.3989}}].

\bibitem{Khachatryan:2016zqb}
{\scshape CMS} collaboration, V.~Khachatryan et~al., \emph{{Search for narrow
  resonances in dilepton mass spectra in proton-proton collisions at $\sqrt{s}$
  = 13 TeV and combination with 8 TeV data}},
  \href{https://doi.org/10.1016/j.physletb.2017.02.010}{\emph{Phys. Lett.}
  {\bfseries B768} (2017) 57}
  [\href{https://arxiv.org/abs/1609.05391}{{\ttfamily 1609.05391}}].

\bibitem{Aaboud:2017yyg}
{\scshape ATLAS} collaboration, M.~Aaboud et~al., \emph{{Search for new
  phenomena in high-mass diphoton final states using 37 fb$^{-1}$ of
  proton--proton collisions collected at $\sqrt{s}=13$ TeV with the ATLAS
  detector}}, \href{https://doi.org/10.1016/j.physletb.2017.10.039}{\emph{Phys.
  Lett.} {\bfseries B775} (2017) 105}
  [\href{https://arxiv.org/abs/1707.04147}{{\ttfamily 1707.04147}}].

\bibitem{Alekhin:2013nda}
S.~Alekhin, J.~Blumlein and S.~Moch, \emph{{The ABM parton distributions tuned
  to LHC data}}, \href{https://doi.org/10.1103/PhysRevD.89.054028}{\emph{Phys.
  Rev.} {\bfseries D89} (2014) 054028}
  [\href{https://arxiv.org/abs/1310.3059}{{\ttfamily 1310.3059}}].

\bibitem{Dulat:2015mca}
S.~Dulat, T.-J. Hou, J.~Gao, M.~Guzzi, J.~Huston, P.~Nadolsky et~al.,
  \emph{{New parton distribution functions from a global analysis of quantum
  chromodynamics}},
  \href{https://doi.org/10.1103/PhysRevD.93.033006}{\emph{Phys. Rev.}
  {\bfseries D93} (2016) 033006}
  [\href{https://arxiv.org/abs/1506.07443}{{\ttfamily 1506.07443}}].

\bibitem{Ball:2017nwa}
{\scshape NNPDF} collaboration, R.~D. Ball et~al., \emph{{Parton distributions
  from high-precision collider data}},
  \href{https://doi.org/10.1140/epjc/s10052-017-5199-5}{\emph{Eur. Phys. J.}
  {\bfseries C77} (2017) 663}
  [\href{https://arxiv.org/abs/1706.00428}{{\ttfamily 1706.00428}}].

\bibitem{Butterworth:2015oua}
J.~Butterworth et~al., \emph{{PDF4LHC recommendations for LHC Run II}},
  \href{https://doi.org/10.1088/0954-3899/43/2/023001}{\emph{J. Phys.}
  {\bfseries G43} (2016) 023001}
  [\href{https://arxiv.org/abs/1510.03865}{{\ttfamily 1510.03865}}].

\bibitem{deFlorian:2014vta}
D.~de~Florian, J.~Mazzitelli, S.~Moch and A.~Vogt, \emph{{Approximate N$^{3}$LO
  Higgs-boson production cross section using physical-kernel constraints}},
  \href{https://doi.org/10.1007/JHEP10(2014)176}{\emph{JHEP} {\bfseries 10}
  (2014) 176} [\href{https://arxiv.org/abs/1408.6277}{{\ttfamily 1408.6277}}].

\bibitem{Das:2020adl}
G.~Das, S.~Moch and A.~Vogt, \emph{{Approximate four-loop QCD corrections to
  the Higgs-boson production cross section}},
  \href{https://arxiv.org/abs/2004.00563}{{\ttfamily 2004.00563}}.

\end{thebibliography}\endgroup
